\documentclass[12pt]{article}                            
\usepackage{amssymb,amsmath}
\begin{document}
\vspace{10mm}
\begin{center}
                                                                        
\vspace{10mm}
\begin{center}
        
{\Large \bf The symmetry, connecting the processes in 2- and 4-dimensional
space-times, and the value $\alpha_0 = 1/4\pi$ for the bare fine structure
constant }\\

\vspace{10mm}  V.I. Ritus ${}^\dag$\\

\vspace{3mm}{$\dag$\it  Lebedev Physical Institute, 119991, Moscow, Russia\\
e-mail: ritus@lpi.ru}

\vspace{2mm}

\end{center}
                
\end{center}
\begin{abstract}
The symmetry manifests itself in exact mathematical relations between the 
Bogoliubov coefficients for the processes induced by accelerated point mirror
in 1+1-space and the current (charge) densities for the processes caused by
accelerated point charge in 3+1-space. The spectrum of pairs of Bose (Fermi)
massless quanta, emitted by mirror, coincides with the spectrum of photons
(scalar quanta), emitted by electric (scalar) charge up to multiplier $e^2/
\hbar c$. The integral connection of the propagator of a pair of oppositely 
directed massless particles in 1+1-space with the propagator of a single 
particle in 3+1-space leads to equality of the vacuum-vacuum amplitudes for 
charge and mirror if the mean number of created particles is small and the 
charge $e=\sqrt{\hbar c}$. Due to the symmetry the mass shifts of electric and 
scalar charges, the sources of Bose-fields with spin 1 and 0 in 3+1-space, for 
the trajectories with subluminal relative velocity $\beta_{12}$ of the ends 
and maximum proper acceleration $w_0$ are expressed in terms of heat capacity 
(or energy) spectral densities of Bose- and Fermi-gases of massless particles 
with temperature $w_0/2\pi$ in 1+1-space. Thus, the acceleration excites the 
one-dimensional oscillations in the proper field of a charges and the energy of 
oscillations is partly deexcited in the form of real quanta and partly remains 
in the field. As a result, the mass shift of accelerated electric charge is
nonzero and negative , while that of scalar charge is zero. The symmetry is 
extended to the mirror and charge interactions with the fields carrying 
spacelike momenta and defining the Bogoliubov coefficients $\alpha^{B,F}$. The 
traces ${\rm tr}\,\alpha^{B,F}$ describe the vector and scalar interactions of 
accelerated mirror with a uniformly moving detector and were found in analytical 
form for two mirror's trajectories with subluminal velocities of the ends.
The symmetry predicts one and the same value $e_0=\sqrt{\hbar c}$ for electric 
and scalar charges in 3+1-space. The arguments are adduced in favour of that 
this value and the corresponding value $\alpha_0=1/4\pi$ for fine structure 
constant are the bare, nonrenormalized values.

\end{abstract}

\section{Introduction}
The Hawking's mechanism for particle production at the black hole formation is
analogous to the emission from an ideal mirror accelerated in vacuum [1].
In its turn there is a close analogy between the radiation of pairs of scalar
(spinor) quanta from accelerated mirror in 1+1 space and the radiation of
photons (scalar quanta) by an accelerated electric (scalar) charge in 3+1 space
[2,3]. Thus all these processes turn out to be mutually related. In problems
with moving mirrors the $in$-set $\phi_{in\,\omega'},\,\phi_{in\,\omega'}^{*}$ 
and $out$-set $\phi_{out\,\omega},\,\phi_{out\,\omega}^{*}$ of the wave equation 
solutions are usually used. For massless scalar field they look as follows:
\begin{equation}
\phi_{in\,\omega'}(u,v)=\frac{1}{\sqrt{2\omega'}}[e^{-i\omega' v}-e^{-i\omega' 
f(u)}],\qquad 
\phi_{out\,\omega}(u,v)=\frac{1}{\sqrt{2\omega}}[e^{-i\omega g(v)}-e^{-i\omega 
u}],
\end{equation}
with zero boundary condition $\phi\vert_{traj}=0$ on the mirror's trajectory.
Here the variables $u=t-x,\,v=t+x$ are used and the mirror's (or charge's)
trajectory on the $u,\,v$ plane is given by any of the two mutually inverse
functions $v^{mir}=f(u),\;u^{mir}=g(v)$. 
                        
For the $in$- and $out$-sets of massless Dirac equation solutions see [3]. Dirac
solutions differ from (1) by the presence of bispinor coefficients at $u$- and
$v$-plane waves. The current densities corresponding to these solutions have 
only tangential components on the boundary. So, the boundary condition both for
scalar and spinor field is purely geometrical, it does not contain any 
dimensional parameters.

The Bogoliubov coefficients $\alpha_{\omega'\omega},\,\beta_{\omega'\,\omega}$
appear as the coefficients of the expansion of the $out$-set solutions in the 
$in$-set solutions; the coefficients $\alpha^*_{\omega'\omega},\,\mp\beta_
{\omega'\omega}$ arise as the coefficients of the inverse expansion. The upper 
and lower signs correspond to scalar (Bose) and spinor (Fermi) field. The 
explicit form of Bogoliubov coefficients is very simple:
\begin{equation}
\alpha^B_{\omega'\omega},\;\beta^{B*}_{\omega'\omega}=
\sqrt{\frac{\omega'}{\omega}}\int_{-\infty}^\infty dv\,e^{i\omega' v\mp i\omega
g(v)}=\pm\sqrt{\frac{\omega}{\omega'}}\int_{-\infty}^\infty du\,e^{\mp i\omega u
+i\omega' f(u)}.
\end{equation}
The coefficients $\alpha^F_{\omega'\omega},\,\beta^{F*}_{\omega'\omega}$ for
Fermi-field differ from these representations by the changes 
$\sqrt{\omega'/\omega}\to \sqrt{g'(v)},\,\pm\sqrt{\omega/\omega'}\to 
\sqrt{f'(u)}$ under the integral signs.

Then the mean number $d\bar n_{\omega}$ of quanta radiated by accelerated
mirror to the right semi-space with frequency $\omega$ and wave vector 
$\omega>0$, and the total mean number $\bar N$ of quanta are given by the 
integrals
\begin{equation}      
d\bar n^{B,F}_{\omega}=\frac{d\omega}{2\pi}\int_0^\infty \frac{d\omega'}{2\pi}
\vert \beta^{B,F}_{\omega'\omega}\vert^2,\qquad 
\bar N^{B,F}=\int\!\!\!\!\int\limits_0^\infty \frac{d\omega d\omega'}{(2\pi)^2}
\vert\beta^{B,F}_{\omega'\omega}\vert^2.
\end{equation}
These expressions do not contain $\hbar$, but their interpretation as mean 
numbers of quanta follows from the second-quantized theory. The second-quantized 
theory allows to construct the all possible amplitudes of many-particle creation, 
annihilation and scattering via Bogoliubov coefficients [4,5,6].
 
At the same time the spectra of photons and scalar quanta emitted by electric
and scalar charges moving along the trajectory $x_{\alpha}(\tau)$ in 3+1 space
are defined by the Fourier transforms of the electric current density 4-vector 
$j_{\alpha}(x)$ and the scalar charge density $\rho(x)$, 
\begin{equation}
j_{\alpha}(k),\,\rho (k)=e\int d\tau\,\lbrace \dot x_{\alpha}(\tau),\,1
\rbrace e^{-ik^{\alpha}x_{\alpha}(\tau)},\quad 
j_{\alpha}(x),\,\rho (x)=e\int d\tau\,\lbrace \dot x_{\alpha}(\tau),\,1\rbrace
\delta_4(x-x(\tau)),
\end{equation}
and are given by the formulae
\begin{equation}
d\bar n^{(1,0)}_k=\frac{1}{\hbar c}\lbrace \vert j_{\alpha}(k)\vert ^2,\;
\vert \rho (k)\vert ^2\rbrace\frac{dk_+dk_-}{(4\pi)^2},\qquad
\bar N^{(1,0)}=\frac{1}{\hbar c}\int\!\!\!\!\int\limits_0^\infty \frac{dk_+dk_-}
{(4\pi)^2}\lbrace \vert j_{\alpha}(k)\vert^2,\,\vert \rho (k)\vert^2 \rbrace,
\end{equation}
where the upper index in $d\bar n^{(s)}_k,\;\bar N^{(s)}$, and $k^{\alpha}$ 
denote the spin and 4-momentum of quanta,
$$
k^2=k^2_1+k^2_{\perp}-k^2_0=0,\quad k^2_{\perp}=k^2_0-k^2_1=k_+k_-,\quad
k_{\pm}=k^0\pm k^1,
$$
and in (5) it is supposed that the trajectory $x^{\alpha}(\tau)$ has only $x^0$ 
and $x^1$ nontrivial components, as the mirror's one.

In contrast to quantities in (3), the $d\bar n^{(s)}_k$ and $\bar N^{(s)}$ in
(5) contain $\hbar$ since the charge entering into current and charge densities 
is considered as classical quantity. In essence, the $d\bar n^{(s)}_k$ and 
$\bar N^{(s)}$ can be considered as classical quantities because they are 
obtained from purely classical radiation energy spectrum $d\bar {\cal E}^{(s)}
_k$ divided by the energy $\hbar k^0$ of single quantum, so that
\begin{equation}
d \bar n^{(s)}_k=\frac{d\bar {\cal E}^{(s)}_k}{\hbar k^0},\quad
\bar N^{(s)}=\int \frac{d\bar {\cal E}^{(s)}_k}{\hbar k^0},\quad
k^0=\frac12 (k_+ +k_-).
\end{equation}

The symmetry between the creation of Bose or Fermi pairs by accelerated mirror
in 1+1 space and the emission of single photons or scalar quanta by electric or
scalar charge in 3+1 space consists, first of all, in the coincidence of the
spectra. If one puts $2\omega=k_+,\;2\omega'=k_-$, then
\begin{equation}
\vert\beta^B_{\omega'\omega}\vert^2=\frac{1}{e^2}\vert j_{\alpha}(k_+,k_-)\vert
^2,\qquad \vert\beta^F_{\omega'\omega}\vert^2=\frac{1}{e^2}\vert\rho(k_+,k_-)
\vert^2.
\end{equation}
So, the spectra coincide as a functions of two variables and a functionals of
common trajectory of a mirror and a charge. The distinction in multiplier 
$e^2 /\hbar c$ can be removed if one puts $e^2=\hbar c$.

The symmetry under discussion connecting the classical and quantum theories in
Minkowsky spaces of 4 and 2 dimensions in some sense reminds the duality 
of classical and quantum descriptions in spaces of neighbour dimensions which 
was proposed by G.'t Hooft [7] and L. Susskind [8]. Such a duality really 
discovered by Gubser, Klebanov and Polyakov [9] and by Maldacena [10] for 
different types of quasiclassical supergravity in anti de Sitter space and 
quantum conformal theories on the boundary of this space. It seems plausible 
that the general reason of such a dualities consists in the correspondence 
between a single particle in space of larger dimension and a pair of particles 
in space of smaller dimension. The description of a larger number of particles 
in space of smaller dimension needs in accounting the quantum mechanical 
interference effects.

\section{Symmetry and physical content and distinction of 
$\beta^{*}_{\omega'\omega}$ and $\alpha_{\omega'\omega}$}
                                                                             
It follows from the second-quantized theory that the absolute pair production 
amplitude and the single-particle scattering amplitude are connected by the 
relation 
\begin{equation}
\langle {\rm out}\omega''\omega\vert {\rm in}\rangle=-\sum_{\omega'}
\langle {\rm out}\omega''\vert \omega'{\rm in}\rangle\,\beta^*_{\omega'\omega}.
\end{equation}
It enables to interpret $\beta^*_{\omega'\omega}$ as the amplitude of a source 
of a pair of the massless particles potentially emitted to the right and to the    
left with frequences $\omega$ and $\omega'$ respectively [6]. While the particle
with frequency $\omega$ actually escapes to the right, the particle with
frequency $\omega'$ propagates some time then is reflected by the mirror and
is actually emitted to the right with altered frequency $\omega''$. Then, in the 
time interval between pair creation and reflection of the left particle,
we have the virtual pair with energy $k^0$, momentum $k^1$, and mass $m$:
\begin{equation}
k^0=\omega+\omega',\quad k^1=\omega-\omega',\quad m=\sqrt{-k^2}=2\sqrt{\omega
\omega'}.
\end{equation}

Apart from this polar timelike 2-vector $k^{\alpha}$, very important is the 
axial spacelike 2-vector $q^{\alpha}$,
\begin{equation}
q_{\alpha}=\varepsilon_{\alpha\beta}k^{\beta},\quad q^0=-k^1=-\omega+\omega',
\quad q^1=-k^0=-\omega-\omega'<0.
\end{equation}
With the help of $k^{\alpha}$ and $q^{\alpha}$ the symmetry between $\alpha$
and $\beta$ coefficients becomes clearly expressed:
\begin{equation}
s=1,\qquad e\beta^{B*}_{\omega'\omega}=-\frac{q_{\alpha}j^{\alpha}(k)}
{\sqrt{k_+k_-}},\qquad e\alpha^B_{\omega'\omega}=-\frac{k_{\alpha}j^{\alpha}(q)}
{\sqrt{k_+k_-}},
\end{equation}
\begin{equation}
s=0,\qquad e\beta^{F*}_{\omega'\omega}=\rho(k),\qquad e\alpha^F_{\omega'\omega}
=\rho(q).
\end{equation}

Note that the equations (4) define the current density $j^{\alpha}(k)$
and the charge density $\rho (k)$ as the functionals of the trajectory 
$x^{\alpha}(\tau)$ and the functions of any 2- or 4-vector $k^{\alpha}$. It can
be shown that in 1+1-space $j^{\alpha}(k)$ and $j^{\alpha}(q)$ are the spacelike 
and timelike polar vectors if $k^{\alpha}$ and $q^{\alpha}$ are the timelike 
and spacelike vectors correspondingly.

The boundary condition on the mirror evokes in the vacuum of massless scalar or
spinor field the appearance of vector or scalar disturbance waves bilinear in
massless fields. There are two types of these waves:

1) The waves with amplitude $\alpha_{\omega'\omega}\;(\alpha^*_{\omega'\omega})$
which carry the spacelike momentum directed to the left (right), and

2) The waves with amplitude $\beta^*_{\omega'\omega}\;(\beta_{\omega'\omega})$
which carry the timelike momentum with positive (negative) frequency.

The waves with the spacelike momenta appear even if the mirror is in rest or
moves uniformly (Casimir effect), while the waves with the timelike momenta
appear only in the case of accelerated mirror.

The pair of Bose (Fermi) particles has spin 1 (0) because its source is the
current density vector (charge density scalar), see [11] or the problem 12.15 
in [12].

\section{Vacuum-vacuum amplitude $\langle {\rm out}\vert {\rm in}\rangle =
e^{iW}$, self-action and mass shifts}
It follows from the secondary quantized theory that in the vacuum-vacuum 
amplitude $\langle {\rm out}\vert {\rm in}\rangle = e^{iW}$ the 
${\rm Im}\,W^{B,F}$ is well-defined. According to DeWitt [4], Wald [5] and 
others (including myself [6])
\begin{equation}
2\,{\rm Im}\,W^{B,F}=\pm\frac12{\rm tr}\,\ln(1\pm\beta^+\beta)\quad{\rm or}\quad
\pm{\rm tr}\,\ln(1\pm\beta^+\beta)
\end{equation}
correspondingly to the cases when particle is identical or nonidentical to
antiparticle. We confine ourselves by the last case and by the smallness of the
tr$\,\beta^+\beta\ll 1$. Then
\begin{equation}
2\,{\rm Im}\,W^{B,F}\approx {\rm tr}\,(\beta^+\beta)^{B,F}\equiv\int\!\!\!\!\int
\limits_0^\infty \frac{d\omega d\omega'}{(2\pi)^2}\,\vert \beta^{B,F}_{\omega'
\omega}\vert^2=\bar N^{B,F}.
\end{equation}
By using in the integrand of $\bar N^{B,F}$ the representations (2) for 
$\beta^{B,F}$, the variables $x_{\mp}(\tau)$ and $x_{\pm}(\tau')$ instead of 
$u,\,f(u)$ and $v,\,g(v)$, and hyperbolic variables $\rho,\,\vartheta$ instead 
of $\omega,\,\omega'$, 
\begin{equation}
d\omega d\omega'=\frac12\rho d\rho d\vartheta,\quad \omega=\frac12\rho 
e^{\vartheta},\quad \omega'=\frac12\rho e^{-\vartheta},\quad \rho=
2\sqrt{\omega\omega'},\quad \vartheta=\ln\sqrt{\frac{\omega}{\omega'}},
\end{equation}
one obtains the imaginary part of the causal function in 1+1-space, 
${\rm Im}\,\Delta_2^f(z,\rho)$, after integration over $\vartheta$, and then 
the imaginary part of the causal function in 3+1-space, ${\rm Im}\,\Delta^f_4
(z,\mu)$, after integration over $\rho=m$, the variable which coincides with 
the mass of virtual pair according to (9). This result is a special case of the 
very important integral relation between the causal functions of wave equations 
for $d$- and $d+2$-dimensional space-times [13],
\begin{equation}
\Delta^f_{d+2}(z,\mu)=\frac{1}{4\pi}\int_{\mu^2}^\infty dm^2\,\Delta^f_d (z,m),
\end{equation}
The small mass parameter $\mu=2\sqrt{\omega\omega'}\vert_{min}\ne 0$ is 
introduced instead of zero to avoid the infrared divergency in the following. 
Thus we obtain                                              

\begin{equation}
2\,{\rm Im}\,W^{B,F}={\rm Im}\int\!\!\!\!\int d\tau d\tau'\left\{
\begin{array}{c}
\dot x_{\alpha}(\tau)\dot x^{\alpha}(\tau')\\1
\end{array}
\right\}\Delta^f_4(z,\mu),\qquad z_{\alpha}=x_{\alpha}(\tau)-x_{\alpha}(\tau').
\end{equation}

We may omit the Im-signs from both of sides of this equation and define the
actions for Bose- and Fermi-mirrors in 1+1-space as
\begin{equation}
W^{B,F}=\frac12\int\!\!\!\!\int d\tau d\tau'\left\{
\begin{array}{c}
\dot x_{\alpha}(\tau)\dot x^{\alpha}(\tau')\\1
\end{array}
\right\}\Delta^f_4(z,\mu).
\end{equation}
Compare this with the well known actions for electric and scalar charges in
3+1-space:
\begin{equation}
W_{1,0}=\frac12\,e^2\int\!\!\!\!\int d\tau d\tau'\left\{
\begin{array}{c}
\dot x_{\alpha}(\tau)\dot x^{\alpha}(\tau')\\1
\end{array}
\right\}\Delta^f_4(z,\mu).
\end{equation}
The symmetry would be complete if $e^2=1$, i.e. if the fine structure constant
were $\alpha=1/4\pi$. This "ideal" value of fine structure constant for the
charges would correspond to the ideal, geometrical boundary condition on the 
mirror.

The appearance in action the causal function $\Delta^f_4(z,\mu)$ has a lucid 
physical grounds.

1. The action must represent not only the radiation of real quanta but also the
self-energy and polarization effects. While the first effects are described by
the solutions of homogeneous wave equation the second ones require the
inhomogeneous wave equation solutions which contain information about proper
field of a source. Namely such solutions of homogeneous and inhomogeneous wave
equations are the functions $(1/2)\Delta^1={\rm Im}\,\Delta^f$ and
$\bar \Delta={\rm Re}\,\Delta^f$.

2. While the appearance of ${\rm Im}\,\Delta^f$ in the imaginary part of the 
action (17) is a consequence of mathematical transformation of the integral 
$\bar N^{B,F}$ (similar to the Plancherel theorem), the function 
$\bar \Delta\equiv{\rm Re}\,\Delta^f$ in the real part of the action is unique 
if it appears as the real part of the analytical continuation of the function 
$i\,{\rm Im}\,\Delta^f(z,\mu)$ to negative $z^2$ that is even in $z$ as 
${\rm Im}\,\Delta^f$ itself.
                                                        
Both the propagator $\Delta^f_2(z,m)$ of a virtual pair with mass
$m=\rho=2\sqrt{\omega\omega'}$ in two-dimensional space-time and the mass
spectrum of these pairs arise owing to the transition from the variables 
$\omega,\,\omega'$ to the hyperbolic variables $\rho,\,\vartheta$, which reflect 
the Lorentz symmetry of the problem. Further integration over the mass leads 
to the propagator $\Delta^f_4(z,\mu)$ of a particle moving in four-dimensional 
space-time with the mass $\mu$ equal to the least mass of virtual pairs. Thus, 
the relation (16) is immanent to the Lorentz symmetry and the symmetry, 
connecting the processes in two- and four-dimensional space-times.

For the point-like charges the $W_{1,0}$ contain ultraviolet divergences and 
need in their elimination. The removal of ultraviolet divergences in the 
self-actions $W_{1,0}\vert^F$ of accelerated charges (force $F\ne 0$) consists 
in the subtraction of corresponding self-actions $W_{1,0}\vert^{F=0}$ of 
uniformly moving charges as a result of which the changes 
$$
\varDelta W_{1,0}=W_{1,0}\vert ^F_0=W_{1,0}\vert^{F}-W_{1,0}\vert^{F=0}
$$ 
of self-actions owing to acceleration do not contain ultraviolet singularities, 
have the positive imaginary part, ${\rm Im}\,\varDelta W_{1,0} >0$, and vanish 
together with acceleration.

The following representations for the self-actions of uniformly moving electric 
and scalar charges are very instructive
\begin{equation}
W_{1,0}\vert^{F=0}=\frac 12\,e^2\int\!\!\!\!\int d\tau\,d\tau' \lbrace \dot x_
{\alpha}(\tau)\dot x^{\alpha}(\tau'),\;1 \rbrace\,\Delta^f_4 (z,\mu)\vert^{F=0} 
= \mp\frac{e^2}{4\pi}\cdot \frac{1-i}{2\sqrt{2\varepsilon}}\cdot \tau.  
\end{equation}
They arise if one introduces the integration variable $x=\tau'-\tau$ instead 
of $\tau'$, so that $z^2=-x^2$, puts $\mu=0$, and makes use of representation
$$
\Delta^f_4(z,\mu)\vert_{\mu =0}=-\frac{1}{4\pi^2}\cdot \frac{i}{x^2-
i\varepsilon}=\frac{1}{4\pi^2}(\frac{\varepsilon}{x^4+\varepsilon^2}-
i\frac{x^2}{x^4+\varepsilon^2}),\quad \varepsilon \to 0.
$$
The opposite signs of the self-actions are due to repulsion of like electric 
charges and to attraction of scalar ones. The coefficients before $\tau$ 
are the classical proper energies $-\delta m_{1,0}$ of the charges taken 
with minus sign, and $\sqrt{2\varepsilon}$ characterizes the charge dimension.
Different signs of ${\rm Im}\,W_{1,0}\vert^{F=0}$ lead, according to amplitudes
$\exp (iW_{1,0}\vert^{F=0})$, to disappearance (screening) of electric charge 
and to unlimited growing (antiscreening) of scalar charge.

These extraordinary properties of the self-actions arise from the pointlikeness 
of the charges. For the vector- and scalar-field sources $j^{\alpha}(x)$ and
$\rho (x)$ distributed in space and slow varying in time the self-actions are 
free from singularities and have no imaginary parts [11]:
\begin{equation}
W_{1,0}=\int dt \int \frac{d^3 x\,d^3 x'}{4\pi\vert{\bf x}-{\bf x'}\vert}
\{j_{\alpha}(x)\,j^{\alpha}(x'),\quad \rho (x)\,\rho (x')\}_{t'=t}.
\end{equation}
In this form self-actions contain the Ampere's and Coulomb laws for current and 
charge interactions and the law of attraction of like scalar charges. 
Self-actions (20), (21) are in accordance with general assertion that the 
interaction of like charges transfered by odd-spin quanta leads to repulsion 
while by even-spin quanta - to attraction.

We exemplify here the self-action changes $\varDelta W_{1,0}$ of electric and 
scalar charges due to accelerated motion along the very important 
quasihyperbolic trajectory
\begin{equation}
x(t)=\frac{\beta^2_1}{w_0}-\beta_1\sqrt{\frac{\beta^2_1}{w^2_0}+t^2},\quad
\beta_{1,2}=\pm {\rm th}\,\frac{\theta}{2},\quad
\beta_{12}=\frac{\beta_1-\beta_2}{1-\beta_1\beta_2}={\rm th}\,\theta,
\end{equation}
with initial $\beta_1$ and final $\beta_2$ velocities at $t=\mp \infty$ and
proper acceleration $-w_0$ at $t=0$. The proper acceleration at any moment is
given by the formula
\begin{equation}
a(t)=-\frac{w_0}{(1+t^2/t^2_c)^{3/2}},\quad t_c=\frac{\beta_1}
{w_0\sqrt{1-\beta^2_1}}.
\end{equation}
Therefore, the quasihyperbolic motion is close to hyperbolic one on the time
interval $\vert t \vert < t_c$.

The self-action changes $\varDelta W_{1,0}
(\theta,\lambda)$ are the Lorentz invariant functions of two variables $\theta
={\rm Arth}\,\beta_{12}$ and $\lambda=\mu^2/w^2_0$ with singularities at 
$\lambda=0$ and $\theta=\pm\infty$.

The case $\lambda\to 0,\;\theta$ arbitrary, was considered by author in [13]:
\begin{equation}
\varDelta W_1=\frac{e^2}{8\pi^2}\lbrace \pi (\frac{\theta}{{\rm th}\,\theta}-1)+
i[(\frac{\theta}{{\rm th}\,\theta}-1)\ln\,\frac{4({\rm ch}\,\theta+1)^2}
{\gamma^2\lambda ({\rm ch}\,\theta-1)}+2-\ln 2-{\rm ch}\,\theta\,R(\theta)]
\rbrace,
\end{equation}
\begin{equation}
\varDelta W_0=\frac{e^2}{8\pi^2}\lbrace \pi (1-\frac{\theta}{{\rm sh}\,\theta})+
i[(1-\frac{\theta}{{\rm sh}\,\theta})\ln\,\frac{4({\rm ch}\,\theta+1)^2}
{\gamma^2\lambda ({\rm ch}\,\theta-1)}-2+\ln 2+R(\theta)]\rbrace,
\end{equation}
here $\gamma=1.781,\;R(\theta)$ is even function of $\theta$ related to the 
Euler's dilogarithm $L_2(z)$ [14], 
\begin{equation}
R(\theta)=\int_0^\infty d\alpha\,\frac{\ln ({\rm ch}\,\theta+{\rm ch}\,\alpha)}
{{\rm ch}\,\theta+{\rm ch}\,\alpha}=\frac{L_2 (1-e^{-2\theta})+\theta^2-
\ln 2\cdot\theta}{{\rm sh}\,\theta}.
\end{equation}

For the case $\theta\to \pm\infty,\;\lambda$ arbitrary, considered in [15,16],
\begin{equation}
\varDelta W_{1,0}=-\vert\,\theta \vert \frac{e^2}{8\pi^2}\,S_{1,0}(\lambda),\quad
S_n(\lambda)=(-1)^{n+1}\int_0^\infty dz\,e^{-i\lambda/2z}[e^{iz}\,K_n(iz)-
\sqrt{\frac{\pi}{2iz}}],
\end{equation}
where $K_n(iz)$ is the Mcdonald function. At $\lambda \to 0$
\begin{equation}
S_1(\lambda)=-\pi-i(\ln\frac{4}{\gamma^2\lambda}-1),\quad S_0(\lambda)=-i.
\end{equation}

For the trajectory with subluminal relative velocity $\beta_{12}$ of the ends 
the ${\rm Re}\,\varDelta W_{1,0}$ are given by the formulae
\begin{equation}
{\rm Re}\,\varDelta W_1=\frac{e^2}{8\pi}(\frac{\theta}{{\rm th}\,\theta}-1),\qquad
{\rm Re}\,\varDelta W_0=\frac{e^2}{8\pi}(1-\frac{\theta}{{\rm sh}\,\theta}).
\end{equation}
When $\beta_{12}\to 1$ the trajectory becomes actually hyperbolic one with
charge's velocity $\beta(\tau)=-{\rm th}\,w_0\tau$ at proper time $\tau$, and
$\theta=w_0(\tau_2-\tau_1)\to \infty$. Then
\begin{equation}
{\rm Re}\,\varDelta W_1=\frac{e^2 w_0}{8\pi}\,(\tau_2-\tau_1),\qquad
{\rm Re}\,\varDelta W_0=\frac{e^2}{8\pi},
\end{equation}
while the mass shifts of uniformly accelerated charges are
\begin{equation}
\varDelta m=-\frac{\partial \varDelta W}{\partial \tau_2}=\frac{e^2w_0}{8\pi^2}
\,S(\lambda);\quad {\rm at}\:\lambda\to 0\quad
{\rm Re}\,\varDelta m_1=-\frac{e^2 w_0}{8\pi},\quad {\rm Re}\,\varDelta m_0=0.
\end{equation}

The real parts of the action changes (29) have interesting integral 
representations ascending to Legendre [17]
\begin{equation}
{\rm Re}\,\varDelta W_{1,0}=\frac{e^2}{4\pi}\int_0^\infty dx\,\frac {\sin\,x}
{e^{\pi x/\theta}\mp 1},\qquad \theta={\rm Arth}\,\beta_{12}.
\end{equation}  
If $\beta_{12}$ is close to 1 then on the large interval of the quasihyperbolic 
trajectory the velocity $\beta(\tau)\approx - {\rm th}\,w_0\tau$, i.e. is the 
same as for hyperbolic trajectory, and the parameter $\theta \approx w_0 
(\tau_2-\tau_1)$, where $\varDelta \tau = \tau_2-\tau_1$ is the proper time
interval inside of which the charge moves with acceleration $w_0$ and outside -
with constant initial and final velocities $\beta_1,\;\beta_2$.

On the acceleration interval the mass shift of a charge can be defined by one
of two relations
\begin{equation}
{\rm Re}\,\varDelta m=-\frac{\partial\,{\rm Re}\,\varDelta W}{\partial\,\tau_2},
\qquad  {\rm Re}\,\widetilde{\varDelta m}=-\frac{{\rm Re}\,\varDelta W}
{\varDelta \tau}.
\end{equation}
                
By using the Legendre representation and formula $\theta=w_0(\tau_2-\tau_1)$,
we obtain, according to the first definition,
\begin{gather}
{\rm Re}\,\varDelta m_1 = -\frac{e^2w_0}{8\pi}\,({\rm cth}\,\theta-\frac{\theta}
{{\rm sh^2}\,\theta}),\qquad
{\rm Re}\,\varDelta m_0 = -\frac{e^2w_0}{8\pi}\,\frac{\theta\,{\rm cth}\,\theta
-1}{{\rm sh}\,\theta},   \notag\\
{\rm Re}\,\varDelta m_{1,0}=-e^2\,T\int_0^\infty \frac{d\omega}{2\pi}\,
\frac{\sin\,2\omega \varDelta \tau}{\omega}\,c^{B,F}(\omega/T),\quad
c^{B,F}(z)=\frac{z^2 e^z}{(e^z\mp 1)^2}.
\end{gather}
In the last, spectral representation $T=w_0/2\pi$ is the Davies-Unruh 
"temperature" [18,19], while $c^{B,F}(\omega/T)$ are the heat capacity spectral 
densities of Bose- and Fermi-gases of massless particles in one-dimensional 
space, see Sections 49, 105 in [20].

The ${\rm Re}\,\varDelta m_{1,0}\leqslant 0$ for all finite $\theta \geqslant 0$; 
for $\theta \ll 1$
\begin{equation}
{\rm Re}\,\varDelta m_1=2\,{\rm Re}\,\varDelta m_0 =-\frac{e^2w_0}{8\pi}\,
\frac23\,\theta,
\end{equation}
and for $\theta \to \infty$
\begin{equation}
{\rm Re}\,\varDelta m_1=-\frac{e^2w_0}{8\pi},\qquad 
{\rm Re}\,\varDelta m_0=0.
\end{equation}
Note, that when the duration of acceleration $\varDelta \tau \to \infty$,
the functon
\begin{equation}
\frac{\sin\,2\omega\varDelta \tau}{\omega}\vert_{\varDelta \tau \to \infty}=
\pi\,\delta(\omega).
\end{equation}
The function on the left-hand side is the Fourier-transform of the acceleration
switching function. The acceleration interval can be regulated by the 
scale-changing of the "temperature" parameter and the frequency, $T\to kT,\; 
\omega \to k\omega$, at constant ratio $\omega/T$. Thus, the temperature 
$T=2w_0/\pi$ also can be used [16].
                                        
According to the second definition
\begin{gather}
{\rm Re}\,\widetilde{\varDelta m_1} = -\frac{e^2w_0}{8\pi}\,({\rm cth}\,\theta-
\frac{1}{\theta}),\qquad
{\rm Re}\,\widetilde{\varDelta m_0} = -\frac{e^2w_0}{8\pi}\,(\frac{1}{\theta}-
\frac{1}{{\rm sh}\,\theta}),   \notag\\
{\rm Re}\,\widetilde{\varDelta m_{1,0}}=-e^2\,\int_0^\infty \frac{d\omega}{2\pi}
\,\frac{\sin\,2\omega \varDelta \tau}{\omega}\,\frac{\omega}{e^{\omega/T}\mp 1}.
\end{gather}
In this case the spectral representation contains the energy spectral density
of Bose- or Fermi-gas of massless particles in one-dimensional space. The 
quantities in both representations are connected by the equation
\begin{equation}
{\rm Re}\,\varDelta m_{1,0}= T\frac{\partial}{\partial T}\,{\rm Re}\,\widetilde 
{\varDelta m_{1,0}}
\end{equation}
which is a consequence of the usual relation
\begin{equation}
c^{B,F}(\omega/T)=\frac{\partial}{\partial T}\left(\frac{\omega}{e^{\omega/T}\mp
1}\right)
\end{equation}
between heat capacity and energy, see Sections 14, 42 in [20].
                                                      
As a functions of $\theta$ the shifts ${\rm Re}\,\varDelta m_{1,0}$ and 
${\rm Re}\,\widetilde {\varDelta m_{1,0}} $ differ in magnitude at $\theta 
\lesssim 1$, for example,
\begin{equation}
{\rm Re}\,\varDelta m_{1,0}=2\,{\rm Re}\,\widetilde{\varDelta m_{1,0}},\qquad
\theta \ll 1,
\end{equation}
but have the same limiting values (36) as $\theta \to \infty$. It can be 
understood, for the mass shift formation it is necessary proper time not less 
than inverse acceleration $w^{-1}_0$.
                          
Thus, according to the spectral representations (34), (38) the symmetry being 
discussed reveals itself also in the formation of the mass shifts of electric 
and scalar charges at acceleration. The vector and scalar massless Bose-fields
of the charges in 3+1-space again appear to be connected with the massless
scalar (Bose) and spinor (Fermi) fields in 1+1-space. The symmetry explaines
why the Legendre representations of the self-action changes and mass shifts of 
electric and scalar charges, the sources of the Bose-fields, contain the 
spectral distributions characteristic for Bose- and Fermi-fields in 
one-dimensional space.
                
The symmetry explaines the limiting values (36) of the mass shifts ${\rm Re}\,
\varDelta m_{1,0}$ for uniformly accelerated electric and scalar charges in
3+1-space by the nonzero and zero low-frequency limits of the heat capacity
(or energy) spectral densities for Bose and Fermi gases in 1+1-space:
\begin{equation}
c^{B,F}(\omega/T)\vert_{\omega=0}=1,\;0;\qquad u^{B,F}(\omega)=\frac{\omega}
{e^{\omega/T}\mp 1}\vert_{\omega=0}=T,\;0.
\end{equation}

The appearance of the heat quantum-mechanical distributions in the spectral
representations of the dynamical mass shifts $\varDelta m_{1,0}$ is no less 
intriguing than their appearance in the Hawking effect [1], especially when the
absence of the horizons for the quasihyperbolic trajectory is taken into 
account.

According to spectral formulae for $\varDelta m$ and $\widetilde{\varDelta m}$
the proper field energy of the charges decreases at the acceleration due to 
radiation on the frequencies
\begin{equation}
\omega_n =\frac{\pi (n+1/2)}{2\,\varDelta \tau}
\end{equation}
with even $n=0,\,2,\,\ldots\,$, and increases due to excitation on the frequencies
$\omega_n$ with odd $n=1,\,3,\,\ldots\,$. We can say that the proper field
releases (deconfines) the excitations with even $n$ and confines the ones with 
odd $n$. The rescaling of $T$ and $\omega$ does not change this assertion.
Eventually, for every finite $\theta > 0$ the radiation-excitation balance 
produces the ${\rm Re}\,\varDelta m < 0$, and the ${\rm Re}\,\varDelta W > 0,\;
{\rm Im}\,\varDelta W > 0$.
                                                        
Simultaneous radiation and excitation of the proper field of a charges at 
acceleration is supported by the positive and negative contributions with even
and odd $n$ frequencies $\omega_n$ to the imaginary part of the self-action 
change,
\begin{equation}
{\rm Im}\,\varDelta W=\frac{1}{\pi}{\rm Re}\,\varDelta W\cdot\ln\frac{1}
{\lambda}+\cdots \,,
\end{equation}
more precisely, to its principal, infrared part, see (24),(25).

Due to the symmetry the quantities $\varDelta W^{B,F},\,\varDelta m^{B,F}$ for 
the mirror interacting with massless Bose- or Fermi-field can be obtained from
$\varDelta W_{1,0},\,\varDelta m_{1,0}$ by the change $e^2 \to \hbar c$.

\section{Arguments in favour of the value $\alpha_0 = 1/4\pi$ for the bare fine
structure constant}
At the collisions of a charged particles, for example, two electrons, the 
emission of soft photons takes place, which does not affect the motion of the
colliding charges. As a result, the cross-section of the particle scattering 
with the emission of $n$ soft photons is represented by the formula (98.21) of 
[23]:
\begin{equation}
d\sigma=d\sigma_{scat}\,w(n),\quad w(n)=\frac{\bar n^n}{n!}\,e^{-\bar n},
\end{equation}
where $w(n)$ is the probability of emission of $n$ soft photons in appropriate 
frequency interval $(\omega_1,\omega_2)$ and $\bar n$ is their mean number, 
which can be found by classical electrodynamics. In this paper the vacuum-vacuum 
amplitude is considered which modulus squared is equal to $w(0)=e^{-\bar n}$.

It is important that the main, logarithmic term of $\bar n$, 
\begin{equation}
\bar n=\alpha\,\frac{2}{\pi}\,(\theta\,{\rm cth}\,\theta-1)(\ln\frac{\omega_2}
{\omega_1}+f(\theta)),
\end{equation}
(see Sections 98, 120 in [23], and formula (24) in this paper where 
$2\,{\rm Im}\,\varDelta W_1=\bar n,\,\mu=\omega_1,\,w_0=\omega_2$),
does not depend on the details of the charge motion and is defined by the 
invariant momentum transfer $\xi =q/2m=\sqrt{t}/2m={\rm sh}\,(\theta /2)$,
which together with the total energy $\sqrt{-s}$ defines the main hard process.
Thus, independently of the charge's motion ("trajectory") inside the forming 
region of the hard process, the mean number of photons emitted by the charge 
is defined by the global parameter -- the momentum transfer or the Lorentz 
invariant velocity change $\beta_{12}$ of a charge in the mentioned region,
$\theta={\rm Arth}\,\beta_{12}$.

The $w(0)=e^{-\bar n}$ with the main, logarithmic term for $\bar n$ is given by
Abrikosov formula (136.11) in [23] for high energy and momentum transfer . It 
coincides with (46) where $\omega_1=\omega_m,\,\omega_2=\varepsilon $. In 
Abrikosov approximation the effective (running) fine structure constant 
$\alpha_{eff}(q^2)$ [22,23] does not differ from $\alpha$,
\begin{equation}
\alpha_{eff}(q^2)=\frac{\alpha}{1-(\alpha/3\pi)\,N_i\,\ln\,(q^2/m^2_i)}.
\end{equation}
Here $m_i$ and $N_i$ are the masses and numbers of different type vacuum charges
screening the bare charge, $m_i<q$. For super high momentum transfers this 
formula does not work.

If the variant (b) of the Gell-Mann and Low paper [21] is realized in quantum
electrodynamics, then on the distances less than some supersmall $\Lambda^{-1}
\ll m^{-1}$ the QED is characterized by the finite point bare charge $e_0$ and
the charge density $e_0\,\delta\,(\bf x)$. In more detail, if the bare fine 
structure constant $\alpha_0=e^2_0/4\pi\hbar c$ is finite, then [21]
                  
1) it does not depend on the value of the fine structure constant $\alpha$,

2) the $\alpha$ must be less than $\alpha_0$, and

3) the charge density at very small distances reduces to the delta-function
$e_0 \delta (\bf x)$.

Therefore, at a collision of a charges with total energy $\sqrt{-s}=2E$ and
momentum transfer $\sqrt{t}\approx 2E \gg \Lambda$ the cross-section $d\sigma
_{scat}$ will be defined by the bare charge $e_0$, and $\bar n$ will be given
by the formula
\begin{equation}
\bar n=\alpha_0\,\frac{2}{\pi}\,(\theta\,{\rm cth}\,\theta-1)(\ln\frac{\omega_2}
{\omega_1}+f(\theta)),
\end{equation}
if the frequencies $\omega_1,\,\omega_2$ satisfy the condition $\Lambda\lesssim
\omega_1\ll\omega_2\ll E$. In this case the photon emission does not influence
as before on the dynamics of the hard process though comes from the super small 
region $\sim \Lambda^{-1}$ where the charge is pointlike and equals $e_0$.
Under these conditions the motion of each colliding charge is one-dimensional 
and can be approximate by classical trajectory with fixed $\theta={\rm Arth \,
\beta_{12}}$ related to this super small region.

The symmetry discussed consists in the coincidence of the number spectrum of pairs of Bose
(Fermi) massless quanta, emitted by point mirror in 1+1-space, with the number 
spectrum of photons (massless scalar quanta), emitted by point electric (scalar)
charge. The first one is obtained by quantum field theory with the corresponding
zero boundary condition on the mirror, while the second is obtained by the 
division of classical energy spectrum on $\hbar\omega$. The corresponding 
spectra coincide as a functions of two variables and a functionals of the any 
common trajectory of the mirror and the charge. The only distinction in 
multiplier $e^2/\hbar c$ can be removed if one puts $e^2=\hbar c$.
      
This symmetry is a consequence of

1) the invariant structure of scalar products in quantum theory of scalar and
spinor fields,

2) the pointness of the mirror and the charge,

3) the unaffectness of the quantum emission on the mirror and charge motions,

4) the space one-dimensionality of the motion.
                                
The 2-dimensional model of QFT with a point mirror interacting with the 
secondary quantized Bose (Fermi)-massless field [4] is pure geometrical: it has 
no mass-dimensional parameters and its Planck constant is dimensionless and 
equals 1. The usual Planck constant appears in the comparison of this QFT-model 
results with the results of QED where there are charge, mass, and momenta and 
energies instead of wave vectors and frequencies, or with the results of classical 
electrodynamics where there are charge, mass and the energy of radiation.

The dimensionless multiplier $e^2/\hbar c$, by which the number spectrum of
soft photons ($\hbar \omega \ll mc^2$ in proper system of a charge) in QED 
differs from the number spectrum of Bose-pairs in 2-dimensional QFT-model, is
less than 1 because the charge in QED has finite dimensions $\sim \hbar/mc$ due
to the screening, while the mirror, the source of Bose-pairs, is point-like.
   
If for the superhigh energy and momentum transfer QED has finite charge $e_0$
of vanishingly small dimensions, then these dimensions cannot be defined better 
than inverse energy $\hbar c/\sqrt{-s}$ of two head-on colliding charges. 
Therefore, it is reasonable to assume that at $\sqrt{-s}\approx \sqrt{t}\to
\infty$ the spectrum of photons with frequencies $\Lambda\lesssim \hbar\omega\ll
\sqrt{-s}$ emitted by the bare charge $e_0$ does not differ from the spectrum of 
Bose-pairs radiated by point mirror. Then $e^2_0=\hbar c$ and $\alpha_0=1/4\pi$. 
The Gell-Mann and Low properties of $\alpha_0$ are fulfilled.
    
Let us consider the head-on collision of two electrons with mass $m$, charge 
$e$, and very large energy $E$ on infinity. The elastic 
scattering cross-section depends on two invariants, $s$ and $t$, which in the 
center of mass system are equal to
$$
s=-4E^2,\quad t=2p^2 (1-\cos \vartheta),
$$
where $E=\sqrt{p^2+m^2},\; p$ and $\vartheta$ are the energy, momentum and 
scattering angle of electron in c.m.s. At fixed energy $E$ the smallest distance 
between the charges is attained at the largest momentum transfer, i.e. at
$\vartheta = \pi$, when charges move along the same straight line. Namely in 
this case each of them most deeply penetrates under the screening coat of the
other. Suppose that the total energy is 
enough to penetrate into the region where the electron charges become bare. 
Then the minimal distance between them is equal to
                            
\begin{equation}
r^c_{min}=\frac{\alpha_0}{2E}
\end{equation}
according to classical theory.
                                                                
But according to quantum mechanics at the distance $r$ between the charges the
uncertainty in their momentum will be not less than $\varDelta p\approx 1/r$. 
It may be thought that the charges can not be on the 
distance less than $r^q_{min}$ at which the momentum uncertainty would lead to
the energy greater than $2E$. Then $2E=2\sqrt{M^2+\varDelta p^2},\;\varDelta p
=\sqrt{E^2-M^2}$, where $M$ is the mass of the bare charge. The Gell-Mann and 
Low's pointness of the bare charge forces one to consider that $M\sim E$. 

For the $r^q_{min}$ we have
\begin{equation}
r^q_{min}=\frac{1}{\varDelta p}=\frac{1}{\sqrt{E^2-M^2}}=\frac{2E}{\alpha_0\,
\sqrt{E^2-M^2}}\,r^c_{min}.
\end{equation}
As the minimal quantum distance is distinctly larger than the classical one, the
turning point can be considered to be defined namely by the $r^q_{min}$. Then
the proper acceleration of the charge at the turning point can be found from the 
equation
\begin{equation}
M\,w_0=\frac{e^2_0}{4\pi\,r^{q\,2}_{min}}=\alpha_0\,(E^2-M^2).
\end{equation}
The quantum motion of the charges is little affected by the emission of photons
with frequencies not greater than $w_0$ as the ratio
\begin{equation}
\frac{w_0}{E}=\alpha_0\,\frac{E^2-M^2}{EM}
\end{equation}
is small if the $\alpha_0$ is small and $M\sim E$. Therefore, for the 
calculation of soft photon emission such motion can be approximated by the 
classical trajectory with acceleration $w_0$ at the turning point. As a result, 
for the $\bar n$ one obtains the formula (48), where $\omega_2=w_0$, and the 
parameter $\theta$
\begin{equation}
\theta=2\,{\rm Arsh}\,\frac{q}{2M},\quad q=\varDelta p=\sqrt{E^2-M^2}.
\end{equation}
At $M\sim E$ the force acting on the charge $e_0$ according to (51) is of the
order of $\alpha_0 M^2$ and is small in comparison both with classical force
$M^2/\alpha_0$, when classical electrodynamics becomes self-contradictory, and
with quantum force $M^2$, when QED needs in quantum corrections, see Sec. 75 in
[24].

About 45 years ago E.P. Wigner remarked that the special relativity is the 
physics of Lorentz transformation, and the quantum mechanics is the physics of 
Fourier transformation. Processes induced by a point mirror in 1+1-space are
described by the simplest relativistic quantum theory, which is incarnated in
Bogoliubov coefficients. They are Lorentz-invariant scalar products reduced to
Fourier transforms of massless scalar and spinor wave equation solutions. They
can be considered as concentrate of genetic information about processes in 3+1-
space.

\section{Self-action changes $\varDelta W_{1,0}$ and traces ${\rm tr}\,
\alpha^{B,F}$}
                                                                                
The basis for the symmetry between the processes induced by the mirror in
two-dimensional and by the charge in four-dimensional space-time is the relation
(11), (12) between the Bogoliubov's coefficients $\beta^{B,F}_{\omega'\omega}$ 
and the current density $j^\alpha (k)$ or charge density $\rho (k)$ depending 
on the timelike momentum $k^{\alpha}$. The squares of these quantities represent 
the spectra of real pairs and particles radiated by accelerated mirror and 
charge.

The symmetry is extended to the selfactions of the mirror and the charge and to 
the corresponding vacuum-vacuum amplitudes, cf. (18) and (19). In essence, it is 
embodied in the integral relation (16) between propagators of a massive pair in 
two-dimensional space and of a single particle in four-dimensional space.

The formula (18) for $W^{B,F}$ was obtained provided that the mean number 
$\bar N^{B,F}$ of pairs created is small and the interference of two or more 
pairs is negligible. In the general case the $W^{B,F}$ is given by the formula 
(13), which can be written also in the form
\begin{equation}
2\,{\rm Im}\,W^{B,F}=\pm{\rm tr}\,\ln (\alpha^{+}\alpha)^{B,F},
\end{equation}
since $\alpha^+\alpha \mp\beta^+\beta=1$, see [4], [6]. As is seen from (13), 
the imaginary part of the action differs from zero and then is positive only if 
$\beta \ne 0$, i.e. if the radiation of real particles is happened indeed.

Formula (54) allows to choose for $W^{B,F}$ the expression
\begin{equation}
W^{B,F}=\pm i\,{\rm tr}\,\ln\,\alpha^{B,F},
\end{equation}
that was called natural by DeWitt [4]. However, this expression is by no means 
unique, the expressions with $ \alpha e^{i\gamma}$ or $\alpha^+$ have the same 
imaginary part. Nevertheless, the formula (55) is interesting as
the definition both the real and imaginary parts of the selfactions $W^{B,F}$ by
means of the Bogoliubov's coefficients $\alpha^{B,F}_{\omega'\omega}$ only,
which, according to the formulae (11), (12), reduce to the current density
$j^\alpha (q)$ or to the charge density $\rho (q)$ dependent on the spacelike 
momentum $q^\alpha$. This means that the field of the corresponding perturbations 
propagates in vacuum together with the mirror, comoves it, and, at the same 
time, it contains the information about the radiation of the real quanta.

Unfortunately, the author failed to find a simple integral representation for
the matrix $\ln\,\alpha$. Nevertheless, if one again assumes that the mean 
number of emitted particles is small, then one may consider $\alpha$, or
$i\alpha$, or $\pm i \alpha^{B,F}$ close to 1. Namely the last phase factor is
most acceptable as will be seen below. Then, expanding the $\ln\,(\pm i\alpha
^{B,F})$ near $\pm i\alpha^{B,F}=1$ and confine ourselves by the first term we 
obtain
\begin{equation}
W^{B,F}=\pm i\,{\rm tr}\,\ln\,(\pm i\alpha^{B,F}) \sim
\pm i\,{\rm tr}\,(\pm i\alpha^{B,F}-1)=-{\rm tr}\,\alpha^{B,F}+\ldots\;.    
\end{equation}
These qualitative arguments allow to state that the functionals ${\rm tr}\,
\alpha^{B,F}$ are similar to the corresponding self-actions with opposite sign
and therefore must have the negative imaginary parts. This is confirmed by the
general examples considered below in which at least the initial or the final
velocity of the mirror is subluminal. 

However, as is shown in the next Section, the above reasoning is very crude.
The exact physical meaning of the ${\rm tr}\,\alpha^{B,F}$ is conveyed by the
formula (99) or (102). As a result, each of the traces represents the mass shift 
of a field, entrained by a mirror at acceleration, multiplied by effective 
proper time of the shift formation. This time is of the order of $w^{-1}_0$.

\section{Invariant structure of Bogoliubov coefficients}

Here, using, as an example, the Bogoliubov coefficients for hyperbolic motion of 
the mirror [25,26],  
\begin{equation}
\alpha_{\omega' \omega}^{B,F}=\frac{2}{\sqrt{\varkappa \varkappa'}}\,e^{i(\frac
{\omega}{\varkappa}+\frac{\omega'}{\varkappa'})}K_{1,0}(2i\sqrt{\frac{\omega 
\omega'}{\varkappa \varkappa'}})=\frac{2}{w_0}\,e^{i\frac{\rho}{w_0}{\rm ch}\,
(\vartheta - \alpha)}K_{1,0}(\frac{i\rho}{w_0}).
\end{equation}
\begin{equation}
\beta_{\omega' \omega}^{B,F*}=(-i)^{1,0}\frac{2}{\sqrt{\varkappa \varkappa'}}\,
e^{i(-\frac{\omega}{\varkappa}+\frac{\omega'}{\varkappa'})}K_{1,0}(2\sqrt{\frac
{\omega \omega'}{\varkappa \varkappa'}})=(-i)^{1,0}\frac{2}{w_0}\,e^{-i\frac
{\rho}{w_0}{\rm sh}\,(\vartheta - \alpha)}K_{1,0}(\frac{\rho}{w_0}).
\end{equation}
we shall consider the invariant properties of the coefficients relative to 
Lorentz transformation and the transformation properties relative to transfer
of the reference origin from one point on the trajectory to another.
   
The Bogoliubov coefficients are the functionals of the trajectory and a 
functions of the frequencies $\omega,\,\omega'$ and parameters $\varkappa,\,
\varkappa'$. The latters characterize the mirror trajectory $u^{mir}=g(v)$ near 
the coordinate origin $u=v=0$ chosen on the trajectory:
\begin{equation}
u^{mir}=g(v)=\frac{1}{\varkappa}\left(\varkappa' v+b(\varkappa' v)^2+
\frac13 c(\varkappa' v)^3+\cdots\right).
\end{equation}
The velocity and proper acceleration of the mirror at the point $u=v=0$ are 
equal to
\begin{equation}
\beta_0=\frac{1-\varkappa'/\varkappa}{1+\varkappa'/\varkappa},\qquad
a_0=-b\sqrt{\varkappa \varkappa'}.
\end{equation}
At the Lorentz transformation with velocity $\beta={\rm th}\,\delta$ parameters
$\varkappa,\,\varkappa'$ are transformed just as frequencies $\omega,\,\omega'$:
\begin{equation}
\tilde \omega=\frac{\omega-\beta\,\omega}{\sqrt{1-\beta^2}}=\omega\,e^{-\delta},
\quad \tilde \omega'=\frac{\omega'+\beta\,\omega'}{\sqrt{1-\beta^2}}=
\omega'\,e^{\delta},
\end{equation}
and the product $\tilde \omega \tilde \omega'=\omega \omega'$ is invariant.
Therefore the frequencies $\omega,\,\omega'$ and parameters $\varkappa,\,
\varkappa'$ can be represented in the form
\begin{equation}
\omega=\sqrt{\omega\omega'}\,e^{\vartheta},\; \omega'=\sqrt{\omega\omega'}\,
e^{-\vartheta};\quad \varkappa=\sqrt{\varkappa\varkappa'}\,e^{\alpha},\;
\varkappa'=\sqrt{\varkappa\varkappa'}\,e^{-\alpha}.
\end{equation}
In the coordinate system with velocity $\beta_C$ relative to the laboratory
system
\begin{equation}
\beta=\beta_C=\frac{\omega-\omega'}{\omega+\omega'}={\rm th}\,\vartheta,\qquad
\vartheta=\ln\,\sqrt{\omega/\omega'},
\end{equation}
the frequencies $\omega$ and $\omega'$ of reflected and incident waves coincide 
and are equal to invariant $\sqrt{\omega\omega'}$ while the vectors
\begin{equation}
k^{\alpha}=(k^1,\,k^0)=(\omega-\omega',\,\omega+\omega'),\quad
q^{\alpha}=(q^1,\,q^0)=(-\omega-\omega',\,-\omega+\omega'),
\end{equation}
have only time and only space components correspondingly:
\begin{equation}
k^{\alpha}_C=(0,\,2\sqrt{\omega\omega'}),\qquad
q^{\alpha}_C=(-2\sqrt{\omega\omega'},\,0).
\end{equation}
The given formulae were used in the coefficients (43), (44) for hyperbolic
trajectory
\begin{equation}
t(\tau)=\frac{{\rm sh}\,(w_0 \tau - \alpha)+ {\rm sh}\,\alpha}{w_0},\qquad
x(\tau)=\frac{{\rm ch}\,\alpha-{\rm ch}\,(w_0 \tau -\alpha)}{w_0},
\end{equation}                                               
for which the proper acceleration is equal to $a_0=-\sqrt{\varkappa\varkappa'}
=-w_0$.
         
The velocity of the mirror on this trajectory at the moment $w_0\tau$ is equal 
to
\begin{equation}
\beta (w_0\tau)=\frac{\dot x (w_0\tau)}{\dot t (w_0\tau)}=-{\rm th}\,(w_0\tau-
\alpha).
\end{equation}
The mirror passes the coordinate origin with velocity $\beta_0=\beta (0)=
{\rm th}\,\alpha$ at the moment $w_0\tau =0$, it passes the turning point at the
moment $w_0\tau =\alpha ,\; \beta (\alpha)=0$, and at the moments $w_0\tau_{1,2}
=\alpha \mp (\vartheta-\alpha)$ before and after the turn its velocities are 
equal to
\begin{equation}
\beta (w_0\tau_{1,2})=\pm{\rm th}\,(\vartheta-\alpha)=\pm \frac{\beta_C -
\beta_0}{1-\beta_C \beta_0}=\pm \beta_{C0}.
\end{equation}
The velocities $\beta_C$ and $\beta_{C0}$ are the velocities of the pair of 
waves $\omega, \,\omega'$ in the laboratory system and in the system moving
relative to the laboratory system with velocity $\beta_0$. This last system
will be called the system of detector which moves with constant velocity 
$\beta_0$ and touches to the mirror at the point $t=x=0$.

Thus, the laboratory time intervals
\begin{equation}
\varDelta t_{1,2}=t(w_0\tau_{1,2})-t(\alpha)=\mp \frac{{\rm sh}\,(\vartheta-
\alpha)}{w_0}
\end{equation}
and the laboratory space intervals
\begin{equation}
\varDelta x_{1,2}=x(w_0\tau_{1,2})-x(\alpha)=- \frac{{\rm ch}\,(\vartheta-
\alpha)-1}{w_0}   
\end{equation}
marked off from the turning point define the time and length of deceleration
($w_0\tau_1 \leqslant w_0\tau \leqslant \alpha$) and acceleration
($\alpha \leqslant w_0\tau \leqslant w_0\tau_2$) intervals on the world 
trajectory of the mirror where its velocity changes monotonously in the 
interval
\begin{equation}
-\beta_{C0} \leqslant \beta \leqslant \beta_{C0}
\end{equation}
between opposite in sign values (60) and takes zero value at the turning point.
It is supposed that $\vartheta > \alpha$. In the case $\vartheta < \alpha$ the
moment $\tau_2 < \tau_1$ and the deceleration and acceleration intervals will be
$w_0\tau_2 \leqslant w_0\tau \leqslant \alpha$ and $\alpha \leqslant w_0\tau 
\leqslant w_0\tau_1$ correspondingly.

Let us show that the intervals $\varDelta t_{1,2}$ and $\varDelta x_{1,2}$
are Lorentz-invariants, i.e. do not depend on transition to another Lorentz 
coordinate system. Let the system $\tilde K$ moves with velocity $\beta=
{\rm th}\,\delta$ relative to the laboratory system $K$. Then the mirror motion 
equations in the system $\tilde K$ accept the form
\begin{equation}
\tilde t(w_0\tau)=\frac{t(w_0\tau)-\beta\,x(w_0\tau)}{\sqrt{1-\beta^2}}=
\frac{{\rm sh}\,(\alpha-\delta)+{\rm sh}\,(w_0\tau-\alpha+\delta)}{w_0},
\end{equation}
\begin{equation}
\tilde x(w_0\tau)=\frac{x(w_0\tau)-\beta\,t(w_0\tau)}{\sqrt{1-\beta^2}}=
\frac{{\rm ch}\,(\alpha-\delta)+{\rm ch}\,(w_0\tau-\alpha+\delta)}{w_0},
\end{equation}
differing from the nontransformed ones by the shift $\alpha \to \tilde \alpha =
\alpha-\delta$ of the parameter $\alpha$.

The velocity of the mirror in new system is 
\begin{equation}
\tilde \beta (w_0\tau)=\frac{\Dot {\Tilde x} (w_0\tau)}{\Dot {\Tilde t}
(w_0\tau)}=-{\rm th}\,(w_0\tau - \alpha + \delta).
\end{equation}
At the moment $w_0\tau = 0$ of passage through the origin the velocity is equal
to $\tilde \beta_0 = \tilde \beta (0) = {\rm th}\,(\alpha - \delta)$; the 
turning point is went through at the moment $w_0\tau = \alpha - \delta$.

Since the frequencies $\omega , \,\omega'$ at the Lorentz-transformation with
velocity $\beta = {\rm th}\,\delta$ go over into the frequencies $\tilde 
\omega ,\,\tilde \omega'$,
\begin{equation}
\tilde \omega = \sqrt{\omega \omega'}\,e^{\vartheta-\delta}, \quad                                                
\tilde \omega' = \sqrt{\omega \omega'}\,e^{-\vartheta+\delta},                                                 
\end{equation}
and differ from nontransformed ones by the shift $\vartheta \to \tilde \vartheta 
= \vartheta -\delta$ of the parameter $\vartheta$,
the velocity $\beta_C = {\rm th}\,\vartheta$ of the pair of waves $\omega ,\,
\omega'$ goes into the velocity $\tilde \beta_C = {\rm th}\,(\vartheta-\delta)$
of the Lorentz-transformed pair of waves $\tilde \omega ,\,\tilde \omega'$.
However, the relative velocity of this pair of waves and detector,
\begin{equation}
\tilde \beta_{C0}=\frac{\tilde \beta_C - \tilde \beta_0}{1-\tilde \beta_C\tilde
\beta_0}={\rm th}\,(\vartheta-\alpha)=\beta_{C0},
\end{equation}
remains unchanges because $\tilde \vartheta-\tilde \alpha = \vartheta-\alpha$,
see (62).

In the new system the time and length of the intervals of deceleration
$(w_0\tilde \tau_1=2\alpha-\vartheta-\delta\leqslant w_0\tau \leqslant \alpha-
\delta)$ and acceleration ($\alpha-\delta\leqslant w_0\tau \leqslant 
w_0\tilde \tau_2=\vartheta-\delta$) from the same initial velocity $\tilde \beta
(w_0\tilde \tau_1)={\rm th}\,(\vartheta-\alpha)$ to the same final velocity
$\tilde \beta (w_0\tilde \tau_2)=-{\rm th}\,(\vartheta-\alpha)$ do not depend on 
parameter $\delta$ and remain the former functions of the Lorentz-invariant
difference $\vartheta-\alpha=\tilde \vartheta-\tilde \alpha$:
\begin{equation}
\varDelta \tilde t_{1,2}=\tilde t (w_0\tilde \tau_{1,2})-\tilde t (\alpha-
\delta)=\mp \frac{{\rm sh}\,(\vartheta-\alpha)}{w_0}=\varDelta t_{1,2},
\end{equation}
\begin{equation}
\varDelta \tilde x_{1,2}=\tilde x (w_0\tilde \tau_{1,2})-\tilde x (\alpha-
\delta)=- \frac{{\rm ch}\,(\vartheta-\alpha)-1}{w_0}=\varDelta x_{1,2}. 
\end{equation}
This difference is nothing but the proper time (multiplied by $w_0$) of the 
mentioned deceleration or acceleration.

For the parameter $\delta=\alpha$ the system $\tilde K$ moves with velocity
$\beta_0$ relative to the laboratory system and coincides with the proper system
of the detector which the mirror touches to at its turning point $\tilde t=
\tilde x=0$. The frequencies $\tilde \omega,\,\tilde \omega'$ of the waves of
pair in the detector system will be denoted as $\Omega,\,\Omega'$:
\begin{equation}
\Omega=\omega\,\sqrt{\frac{\varkappa'}{\varkappa}},\qquad
\Omega'=\omega'\,\sqrt{\frac{\varkappa}{\varkappa'}}.
\end{equation}
Evidently, they are Lorentz-invariant quantities.

In this system $\tilde \beta_0=0$, and the invariant relative velocity
\begin{equation}
\beta_{C0}=\tilde \beta_{C0}=\tilde \beta_C=\frac{\Omega-\Omega'}{\Omega+
\Omega'}={\rm th}\,\varTheta={\rm th}\,(\vartheta-\alpha),\quad
\varTheta=\ln\,\sqrt{\frac\Omega\Omega'}=\ln\,\sqrt{\frac{\omega\varkappa'}
{\omega'\varkappa}}=\vartheta-\alpha,               
\end{equation}
coincides with the velocity $\tilde \beta_C$ of pair of waves $\Omega,\,\Omega'$
and is defined by the ratio $\Omega/\Omega'$ of the transformed frequencies only.

The intervals $\varDelta \tilde t_{1,2},\,\varDelta \tilde x_{1,2}$ are given by 
the formulae (77,78), where $\delta=\alpha,\,w_0\tilde \tau_{1,2}=\mp (\vartheta-
\alpha)$ and $\tilde t(0)=\tilde x(0)=0$. Therefore,
\begin{equation}
\varDelta \tilde t_{1,2}=\tilde t (w_0\tilde \tau_{1,2})=
\mp \frac{{\rm sh}\,(\vartheta-\alpha)}{w_0}=\varDelta t_{1,2},
\end{equation}
\begin{equation}
\varDelta \tilde x_{1,2}=\tilde x (w_0\tilde \tau_{1,2})=
- \frac{{\rm ch}\,(\vartheta-\alpha)-1}{w_0}=\varDelta x_{1,2}. 
\end{equation}

At the switching off the acceleration the trajectory of the mirror coincides
with the trajectory of the detector and the $\alpha_{\omega'\omega}$ becomes the
matrix diagonal in frequencies (79):
\begin{equation}
\alpha^{B,F}_{\omega'\omega}=2\pi\,\delta(\Omega-\Omega').
\end{equation}
Its functional dependence on the trajectory reduces in this case to the 
dependence on the parameter $\beta_0={\rm th}\,\alpha=
{\rm th}\,(\ln \sqrt{\varkappa/\varkappa'})$ or the Doppler factor $\sqrt
{\varkappa/\varkappa'}$ entering into $\Omega,\,\Omega'$.
           
In the absence of acceleration the frequencies $\omega,\,\omega'$ satisfy the 
condition $\Omega=\Omega'$, and the velocities $\beta_C$ and $\beta_0$ coincide.
Acceleration leads to nonzero Bogoliubov coefficients $\beta_{\omega'\omega}
\ne 0$ and to the absence of the connection $\Omega=\Omega'$ or $\beta_C=
\beta_0$. Distinction between the frequencies $\Omega,\,\Omega'$ or velocities 
$\beta_C,\,\beta_0$ can be described by the invariant relative velocity 
$\beta_{C0}$, see (68) and (76), and leads to the appearance of invariant 
phases of Bogoliubov coefficients defined by this parameter.

With the help of intervals (69,70) the Bogoliubov coefficients can be written in the
form
\begin{equation}
\alpha^{B,F}_{\omega'\omega}=\frac{2}{w_0}\,e^{-i\rho \varDelta x_2+i\rho/w_0}
K_{1,0}(\frac{i\rho}{w_0}),\qquad
\beta^{B,F}_{\omega'\omega}=\frac{2(-i)^{1,0}}{w_0}\,e^{-i\rho \varDelta t_2}
K_{1,0}(\frac{\rho}{w_0}),
\end{equation}
i.e. in the form of proper functions of invariant operators $-i\partial/
\partial\varDelta x_2$ and $i\partial/\partial\varDelta t_2$:
\begin{equation}
-i\frac{\partial\alpha}{\partial\varDelta x_2}=-\rho\,\alpha,\qquad
i\frac{\partial\beta^*}{\partial\varDelta t_2}=\rho\,\beta^*,
\end{equation}
with invariant proper values of momentum transfer $-\rho$ and mass $\rho$
correspondingly.

Thus the phases of the coefficients (84) are defined by the length $\varDelta 
x_{1,2}$ or the time $\varDelta t_{1,2}$ of mirror motion near the turning 
point where the velocity of the mirror changes its sign and does not exceed in 
magnitude the velocity of pair created with time-like momentum.

In one and the same laboratory system it can be introduced two coordinate 
systems $K$ and $K'$, connected by the parallel shift of space-time 
coordinates
\begin{equation}
x=x_1+x',\qquad t=t_1+t'.
\end{equation}
Monochromatic in- and out-waves in the $K$ and $K'$ systems are differed only by
phase multipliers:
\begin{equation}
e^{-i\omega' v}=e^{-i\omega' v_1}\cdot e^{-i\omega' v'},\qquad
e^{-i\omega u}=e^{-i\omega u_1}\cdot e^{-i\omega u'}.
\end{equation}
Therefore, the Bogoliubov coefficients in the systems $K$ and $K'$ are also 
differed by phase multipliers:
\begin{gather}
\alpha_{\omega'\omega}=e^{-i(q\varDelta)}\cdot \alpha'_{\omega'\omega},\quad
-(q\varDelta)=\omega' v_1-\omega u_1,  \notag \\
\beta^*_{\omega'\omega}=e^{-i(k\varDelta)}\cdot \beta'^*_{\omega'\omega},\quad
-(k\varDelta)=\omega' v_1+\omega u_1,
\end{gather}
Here $\varDelta^{\alpha}=(x_1, t_1)$ is 2-vector of the shift, and $k^{\alpha}$
and $q^{\alpha}$ are the wave 2-vectors (9,10).

Particularly, the origin $x=t=0$ of coordinate system $K$ can be chosen at the 
point of the trajectory, where the mirror has nonzero velocity $\beta_0$, and 
the origin $x'=t'=0$ of the coordinate system $K'$ - at the turning point, where
$\beta_1=0$. Then $x_1, t_1$ are the coordinates of the turning point in the 
$K$-system. In this case for the hyperbolic trajectory we have
\begin{equation}
u_1=\frac{1}{w_0}-\frac{1}{\varkappa},\quad v_1=\frac{1}{\varkappa'}-\frac{1}
{w_0};\qquad w_0=\sqrt{\varkappa\varkappa'},\,\beta_0={\rm th}\,\alpha ,\,
\alpha=\ln\,\sqrt{\frac{\varkappa}{\varkappa'}}.
\end{equation}
And the phases of the corresponding multipliers in (88) are equal to the 
differences of phases of Bogoliubov coefficients (57,58) with nonzero and zero 
values of the parameter $\alpha$:
\begin{gather}
-(q\varDelta)=\frac{\omega}{\varkappa}+\frac{\omega'}{\varkappa'}-\frac{\omega 
+\omega'}{w_0}=\frac{\rho}{w_0}{\rm ch}\,(\vartheta-\alpha)-\frac{\rho}{w_0}
{\rm ch}\,\vartheta,    \notag \\
-(k\varDelta)=-\frac{\omega}{\varkappa}+\frac{\omega'}{\varkappa'}-\frac{-\omega 
+\omega'}{w_0}=-\frac{\rho}{w_0}{\rm sh}\,(\vartheta-\alpha)+\frac{\rho}{w_0}
{\rm sh}\,\vartheta.
\end{gather}

The phases of Bogoliubov coefficients can be written as scalar products
\begin{equation}
\frac{\omega}{\varkappa}+\frac{\omega'}{\varkappa'}=-(q\varDelta x),\qquad
-\frac{\omega}{\varkappa}+\frac{\omega'}{\varkappa'}=-(k\varDelta x),
\end{equation}
of 2-vectors $q^{\alpha},\,k^{\alpha}$ defined only by the frequencies 
$\omega,\,\omega'$ and spacelike 2-vector $\varDelta x^{\alpha}$ defined only by 
the parameters $\varkappa,\,\varkappa'$:
\begin{equation}
\varDelta x^1=\frac{1}{2\varkappa}+\frac{1}{2\varkappa'},\quad
\varDelta x^0=\frac{1}{2\varkappa'}-\frac{1}{2\varkappa}.
\end{equation}
The length of $\varDelta x^{\alpha}$ is equal to $1/\sqrt{\varkappa\varkappa'}$
that is equal to $1/w_0$ for hyperbolic trajectory.

Consequently, we have the following forms for the phases
\begin{gather}
-\rho (\varDelta x_2 - \frac{1}{w_0})=-(q\varDelta x)=\frac{\rho}{w_0}{\rm ch}\,
(\vartheta-\alpha),\notag \\
-\rho\,\varDelta t_2=-(k\varDelta x)=-\frac{\rho}{w_0}{\rm sh}\,(\vartheta-
\alpha). 
\end{gather}

Vector $\varDelta x^{\alpha}$ is closely related to acceleration 2-vector 
$a^{\alpha}$ which for the trajectory $u^{mir} = g(v)$ is given by the 
expression
\begin{equation}
a^{\alpha}=(a^1,\,a^0)=-\frac{g''}{4g'^2}(1+g',\,1-g'),\qquad g=g(v).
\end{equation}
At the point $u=v=0$ we get
\begin{equation}
a^{\alpha}_0=a_0\frac{\varDelta x^{\alpha}}{\sqrt{\varDelta x^2}},\qquad
a_0=-b\sqrt{\varkappa\varkappa'},
\end{equation}
where $a_0$ is the proper acceleration at zero point.

The Lorentz-invariant ${\rm tr}\,\alpha$ was defined [26] by the formula 
\begin{equation}
{\rm tr}\,\alpha=\int\!\!\!\!\int\limits_0^\infty\frac{d\omega d\omega'}
{(2\pi)^2}\,\alpha_{\omega' \omega}\,2\pi\,\delta\left(\sqrt{\frac{\varkappa'}
{\varkappa}}\omega-\sqrt{\frac{\varkappa}{\varkappa'}}\omega'\right),\qquad
\Omega=\sqrt{\frac{\varkappa'}{\varkappa}}\omega,\qquad
\Omega'=\sqrt{\frac{\varkappa}{\varkappa'}}\omega',
\end{equation}
in which the Lorentz-invariant argument of $\delta$-function is the difference
of the frequencies $\Omega$ and $\Omega'$ of reflected and incident waves in the 
proper system of the mirror at zero point $u=v=0$ where the mirror has velocity 
$\beta_0$ and acceleration $a_0=-b\sqrt{\varkappa\varkappa'}$. The multipliers 
$\sqrt{\varkappa'/\varkappa},\,\sqrt{\varkappa/\varkappa'}$ are the Doppler 
factors connecting the frequencies in the laboratory system and zero point proper 
system of the mirror (or proper system of the detector).

Thus in the trace formation of matrix $\alpha$ its elements diagonal in 
invariant frequencies are involved, i.e. the elements $\alpha_{\omega'\omega}$
where $\omega/\varkappa=\omega'/\varkappa'$. Note that matrix elements $\alpha
_{\omega'\omega},\;\beta^*_{\omega'\omega}$ being the scalar functions of the
frequencies $\omega,\,\omega'$ can be written in the detector system if one 
performs the changes
\begin{gather}
\omega,\,\omega'\to \Omega,\,\Omega';\;u,\,v\to U=\sqrt{\frac{\varkappa}
{\varkappa'}} u,\,V=\sqrt{\frac{\varkappa'}{\varkappa}} v;\;\notag\\
f(u),\,g(v)\to F(U)=\sqrt{\frac{\varkappa'}{\varkappa}} f(u),\,G(V)=
\sqrt{\frac{\varkappa}{\varkappa'}} g(v),
\end{gather}
in their expressions (2). Then
\begin{equation}
\alpha_{\omega'\omega} = A_{\Omega'\Omega},\quad 
\beta^*_{\omega'\omega} = B^*_{\Omega'\Omega},
\end{equation}
and the diagonal elements $A_{\Omega\Omega}$ with $\Omega=\Omega'=\sqrt{\omega
\omega'}$ are involved in the trace (96).

For the trajectories in the Minkowsky plane on the left from their tangent line 
$X^{\alpha}(\tau')$ at zero point the coordinate $z^1=X^1(\tau')-x^1(\tau)
\geqslant 0$. For these trajectories the ${\rm tr}\,\alpha$ can be transformed 
to the form [26]
\begin{equation}
{\rm tr}\,\alpha^{B,F}=\pm i\int\!\!\!\!\int d\tau d\tau'\left\{
\begin{array}{c}
\dot x_{\alpha}(\tau)\dot X^{\alpha}(\tau')\\1
\end{array}
\right\}\Delta^{LR}_4(z,\nu),\qquad z^\alpha=X^\alpha (\tau')-x^\alpha (\tau),
\end{equation}
where the singular function $\Delta^{LR}_4(z,\nu)$ differs from the causal 
function $\Delta_4^f (z,\mu)$ by complex conjugation and the replacement 
$\mu\to i\nu$ (or by the replacement $z^2\to-z^2,\:\mu\to\nu$):  
\begin{equation}
\Delta_4^{LR}(z,\nu)=\frac{1}{4\pi}\delta(z^2)-\frac{\nu}{8\pi\sqrt{z^2}}
\theta(z^2)H^{(2)}_1(\nu \sqrt{z^2})+i\frac{\nu}{4\pi^2\sqrt{-z^2}}\theta(-z^2)
K_1(\nu\sqrt{-z^2}).
\end{equation}

The expression obtained allows to interpret ${\rm tr}\,\alpha^{B,F}$ as a
functional describing the interaction of two vector or scalar sources by means 
of exchange by vector or scalar quanta with spacelike momenta. At the same time
one of the sources moves along the mirror's trajectory while another one moves
along the tangent line to it at zero point. The last source can be considered
as a probe or detector of excitation created by the accelerated mirror in vacuum.

As the detector moves with constant velocity $\beta_0$, its 2-velocity $\dot X^
{\alpha}(\tau')$ does not depend on $\tau'$. Consequently, $\dot x_{\alpha}
(\tau)\dot X^{\alpha}(\tau') = -\gamma_* (\tau)$ is the relative Lorentz-factor 
defined by the relative velocity $\beta_* (\tau)$ of the mirror and detector:
\begin{equation}
\gamma_* (\tau) = \frac{1-\beta (\tau)\,\beta_0}{\sqrt{1-\beta^2(\tau)}\,\sqrt
{1-\beta^2_0}} = \frac{1}{\sqrt{1-\beta^2_*(\tau)}},\qquad \beta_*(\tau)=
\frac{\beta (\tau) - \beta_0}{1-\beta (\tau)\,\beta_0},
\end{equation}
and is the Lorentz-invariant quantity for each $\tau$. Then
\begin{equation}
{\rm tr}\,\alpha^{B,F}=-i\int d\tau \left\{
\begin{array}{c}
\gamma_* (\tau)\\1
\end{array}
\right\}J(\tau,\nu),\qquad
J(\tau ,\nu)=\int d\tau'\,\Delta^{LR}_4 (z(\tau ,\tau'),\nu).
\end{equation}
It is seen from this representation that at $\theta \ne \infty$, when Lorentz-
factor $\gamma_*(\tau)$ is confined on the whole trajectory, the both traces
have the same qualitative behaviour when parameter $\nu \to 0$. It is clear that
their infrared (logarithmic) singularities in this parameter are indebted to
the behaviour of the integral $J(\tau,\nu)$ at $\tau \to \pm \infty$. For the 
trajectories with subluminal relative velocities $\beta_{10},\:\beta_{20}$ of 
the ends both ${\rm tr}\,\alpha^{B,F}$ have infrared singularities at $\nu = 0$. 
Besides, the singularities of ${\rm tr}\,\alpha^B$ differ from those of 
${\rm tr}\,\alpha^F$ only by the values of the relative Lorentz-factor 
$\gamma_* (\tau)$ for initial and final ends of the trajectory, i.e. by the 
factors $1/\sqrt{1-\beta^2_{10}}$ and $1/\sqrt{1-\beta^2_{20}}$. Since the 
infrared singularities from the initial and final ends appear in 
${\rm tr}\,\alpha^F$ with the factors
\begin{equation}
\frac{\sqrt{1-\beta^2_{10}}}{2\beta_{10}},\qquad \frac{\sqrt{1-\beta^2_{20}}}
{2\vert \beta_{20}\vert},
\end{equation}
they disappear in ${\rm tr}\,\alpha^F$ for the trajectories with luminal 
velocities of the ends, $\beta_{10}=1, \:\beta_{20}=-1$, but remain in 
${\rm tr}\,\alpha^B$. The disappearance of singularities in ${\rm tr}\,\alpha^F$ 
for the such trajectories means that the function $J(\tau,\,\nu)$ is integrable 
in $\tau$ at $\tau \to \pm \infty$ even if $\nu = 0$. At the same time the 
function $\gamma_* (\tau)\,J(\tau,\nu)$ is integrable in this region only at 
$\nu \ne 0$.
                                                
The weakening of interaction of scalar charges with increasing their relative 
velocity, contrary to the constancy of interaction of electric charges, is 
connected with different geometrical structure of scalar and vector field 
sources $\rho (x)$ and $j^{\alpha}(x)$. They are given by (4) for pointlike 
charges moving along the trajectory $x^{\alpha}(\tau)$. 

The charges of the scalar and vector field sources are defined by 
the space integrals of their charge densities $\rho ({\bf x},t)$ and 
$j^0({\bf x},t)$, and for pointlike sources equal to
\begin{equation}
Q_0,\;Q_1=\int d^3 x\,\lbrace \rho ({\bf x},t),\;j^0 ({\bf x},t)\rbrace = 
e \int d\tau \lbrace 1,\;\dot x^0(\tau)\rbrace\,\delta (t-x^0(\tau)) =
e\lbrace \gamma^{-1}(t),\;1\rbrace,
\end{equation}
since $d\tau/dt'= \gamma^{-1}(t')$ if $t'=x^0(\tau)$. As is obvious, the  
charge for the pointlike source $T^{\alpha \beta}(x)$ of a tensor field with 
spin 2 increases as the particle's energy, $Q_2=e\gamma (t)$.
                                                                
As is seen from regularized representation
\begin{equation}
{\rm tr}\,\alpha^{B,F}\vert_{reg}=\frac{1}{2\pi}\int_0^\infty ds\lbrack \int_
{-\infty}^\infty dx\,\{1,\,\sqrt{G'(x)}\}\,e^{-is(G(x)-x)}-
\sqrt{\frac{\pi}{ibs}}\rbrack ,\quad s=\frac{\omega}{\varkappa},
\end{equation}
obtained in [26], the ultraviolet divergences in ${\rm tr}\,\alpha^{B,F}$ are
removed by subtraction from the integrand of the first term its asymptotical
expansion in $s$, as $s\to \infty$. The invariant variable $s=\omega/\varkappa=
\sqrt{\omega\omega'/\varkappa\varkappa'}=b\rho/2w_0$ is proportional to momentum 
transfer $\rho$ in units of proper acceleration $w_0$ of the mirror at the point
of its tangency with detector. The subtracted term, being integrated over $\rho$ 
up to large but finite $\rho_{max}$,
\begin{equation}
\frac{1}{2\pi}\int_0^{s_{max}}ds\,\sqrt{\frac{\pi}{ibs}} =\frac{1}{2\pi}
\sqrt{\frac{\pi\rho_{max}}{w_0}}(1-i),
\end{equation}
is one and the same for Bose and Fermi cases and explicitly depends on 
acceleration.

When the space interval $\varDelta x$ between the mirror and detector becomes 
less than $\hbar/2\varDelta p$, the uncontrolled momentum transfer between 
them becomes greater than $\varDelta p$ and leads to ultraviolet divergency in 
nonregularized ${\rm tr}\,\alpha^{B,F}$. As the mirror coordinate near the 
point of tangency with detector changes in time according to the law $x(t)=
-w_0\,t^2/2$, the time interval $\tau$ necessary for the momentum transfer 
$\varDelta p$ is of the order of $\tau \sim 2\sqrt{\hbar/\varDelta p\,w_0}=
2/\sqrt {w_0\rho_{max}}$ if one sets $\varDelta p=\hbar \rho_{max}$. Then the 
subtracted term which regularizes the ${\rm tr}\,\alpha^{B,F}$ acquires the form
\begin{equation}
\frac{1}{2\pi}\sqrt{\frac{\pi\rho_{max}}{w_0}}(1-i)=\frac{1}{4\pi}\sqrt{\pi}
\rho_{max}\,(1-i)\cdot \tau, \qquad \tau \sim 2/\sqrt{w_0\rho_{max}}.
\end{equation}
As distinct from (20), this term has one and the same sign for Bose and Fermi 
cases. This can be understood as a consequence of positive momentum transfer 
from detector to mirror in both cases. The differences in meanings of $\rho_
{max}\sim 1/\sqrt{2\varepsilon}$ and $\tau$ are more understandable.

Unlike $\varDelta W_{1,0}$, describing the change of selfaction of a 
charges due to acceleration, the functionals ${\rm tr}\,\alpha^{B,F}$ describe 
the interaction of accelerated mirror with the probe executing uniform motion 
along the tangent to the mirror's trajectory at the point where mirror has 
acceleration $w_0$. This interaction is transmitted by the vector or scalar 
perturbations created by the mirror in the vacuum of Bose- or Fermi-field and 
carrying the spacelike momentum of the order of $w_0$. According to (100), 
at distances of the order of $w^{-1}_0$ from the mirror, the field of these 
perturbations decreases exponentially in timelike directions and oscillates 
with damped amplitude in spacelike directions. It can be said that such a field 
moves together with the mirror and is its "proper field".  Hence, the probe 
interacts with the mirror for a time of the order of $w^{-1}_0$ while the charge 
all the time interacts with itself and feels the change of interaction over the 
all time of acceleration. Therefore, it is not surprising that the 
$-{\rm tr}\,\alpha^{B,F}$ coincide in essence with $\varDelta W_{1,0}$ if in 
these latter one puts $\tau_2-\tau_1=2\pi/w_0,\;e^2=1$. In other words, the 
${\rm tr}\,\alpha^{B,F}$ are the mass shifts of the mirror's proper field 
multiplied by characteristic proper time of their formation.    

\section{Interaction with proper field of the accelerated mirror moving with
subluminal velocity}

The ${\rm tr}\,\alpha$ for the trajectory with subluminal velocities of the 
ends is an invariant function of the relative velocities $\beta_{12},\,
\beta_{10},\,\beta_{20}$ connected by the relation $\beta_{12}=(\beta_{10}-
\beta_{20})/(1-\beta_{10}\beta_{20})$. Let us consider the regularized 
${\rm tr}\,\alpha^{B,F}$ for two important trajectories.

1. Quasihyperbolic trajectory, given by the formula (22), is time-reversed to
itself. Its representation in $u,\,v$-variables is the following
\begin{equation}
u^{mir}=g(v)=v\,{\rm ch}\,\theta - \frac{\beta_1}{w_0}{\rm sh}\,\theta +
{\rm sh}\,\theta \sqrt{(v-\frac{\beta^2_1}{w_0})^2+a^2},\quad
a=\frac{\beta_1\sqrt{1-\beta^2_1}}{w_0},\quad 
\beta_1={\rm th}\,\frac{\theta}{2}.
\end{equation}
The initial $\beta_1$ and final $\beta_2=-\beta_1$ velocities are subluminal.

Upon using this expression in the representation (2) and introducing the 
variable $x=v-\beta^2_1/w_0$ we obtain
\begin{equation}
\alpha^B_{\omega'\omega} = 2\sqrt{\frac{\omega'}{\omega}}\,e^{i\frac{\omega+
\omega'}{w_0} \beta^2_1}\int_0^\infty dx\,\cos\,[(\omega'-\omega\,{\rm ch}\,
\theta)x]\,e^{-i\omega\,{\rm sh}\,\theta\,\sqrt{x^2+a^2}}.
\end{equation}
According to the formulae 9 and 15 of the section 2.5.25 in [27] this integral
reduces to the modified Bessel and Hankel functions and we finally have                                                                       
\begin{equation}
\alpha^B_{\omega'\omega}=2ia\,{\rm sh}\,\theta\,\sqrt{\frac{\omega \omega'}{Q}}\,
e^{i\frac{\omega+\omega'}{w_0}\beta^2_{10}}\,K_1(a\sqrt{Q}),\quad
-\pi a\,{\rm sh}\,\theta\,\sqrt{\frac{\omega \omega'}{-Q}}\,
e^{i\frac{\omega+\omega'}{w_0}\beta^2_{10}}\,H^{(2)}_1(a\sqrt{-Q}),
\end{equation}
for $Q=\omega^2+\omega'^2-2\,\omega \omega'\,{\rm ch}\,\theta \gtrless 0$. 
As usual, $\theta={\rm Arth}\,\beta_{12} $ is the Lorentz-invariant parameter 
defined by the relative velocity of the ends.

The corresponding Bogoliubov coefficient for Fermi-case is more complicate:                      
\begin{equation}
\alpha^F_{\omega'\omega}=a\,e^{i\frac{\omega+\omega'}{w_0}\beta^2_{10}}
\int_{-\infty}^\infty dt\,\sqrt{{\rm sh^2}t+{\rm ch^2}\frac{\theta}{2}}\,\exp 
[ia((\omega'-\omega)\,{\rm ch}\frac{\theta}{2}\,{\rm sh}t-(\omega'+\omega)\,
{\rm sh}\frac{\theta}{2}\,{\rm ch}t)].
\end{equation}

As the velocity of the mirror at the point $u=v=0$ (and then the detector 
velocity) is equal to zero, $\beta_0=0$, the initial and final velocities 
$\beta_1,\,\beta_2$ can be considered as invariant relative velocities 
$\beta_1=\beta_{10},\,\beta_2=\beta_{20}$ of the mirror and detector at 
$t=\mp \infty$.

According to the definition (105) we obtain
\begin{equation}
{\rm tr}\,\alpha^B\vert_{reg} = \frac{{\rm cth}\,\theta /2}{2\pi}\,[-\frac{\pi}{2}-
i(\ln \frac{2}{\gamma \varepsilon} - 1)],\qquad \varepsilon = \nu /w_0,
\end{equation}

\begin{equation}
{\rm tr}\,\alpha^F\vert_{reg} = \frac{1}{2\pi}\,\lbrace \frac{1}{{\rm sh}\,\theta /2}\,
[-\frac{\pi}{2} - i(\ln \frac{2}{\gamma \varepsilon} - 1)]+ i[{\rm th}\,\frac
{\theta}{2}\,{\rm \bf B}(k) + \frac{\ln\,{\rm ch}\,\theta /2}{{\rm sh}\,
\theta /2}]\rbrace,
\end{equation}
$$
{\rm \bf B}(k)=\int_0^{\pi /2}\,\frac{\cos^2 \varphi \,d\varphi}{\sqrt{1-k^2
\sin^2\,\varphi}}, \quad k={\rm th}\,\frac{\theta}{2}.
$$
Here ${\rm \bf B}(k)$ is one of the elliptic integrals [14].

In both ${\rm tr}\,\alpha^{B,F}$ the infrared singularities were removed by 
introducing the small parameter $\varepsilon$ (the least momentum transfer in 
$w_0$ units), while the ultraviolet singularities were eliminated as was written 
above.

The function
$$
R(\theta)={\rm th}\,\frac{\theta}{2}\,{\rm \bf B}(k)+\frac{\ln\,{\rm ch}\,\theta
/2}{{\rm sh}\,\theta/2}
$$  
is equal to zero at $k=0$, grows almost linearly with $k$, reaches maximal value
$\approx 1.28$ at $k \approx 0.97$ and then decays rapidly to 1 as $k \to 1$.

2. The Airy's semiparabola with in-tangent line to inflection point is given by
\begin{gather}
\varkappa u^{{\rm mir}}(v)=(1-b^2/c)\,\varkappa'v-b^3/3 c^2,\qquad
-\infty <v\leqslant v_0, \notag \\
=\varkappa'v+b\,\varkappa'^2 v^2+(1/3)\,c\,\varkappa'^3 v^3,\qquad
v_0\leqslant v<\infty,
\end{gather}
where the inflection point $v_0=-b/\varkappa' c,\;b>0 $, and $c>b^2$ due to the 
timelikeness of the trajectory. The initial velocity is subluminal, but the 
final one is luminal.

By using this trajectory in integral 
representations for Bogoliubov coefficients we find that
\begin{equation}
\alpha^B_{\omega'\omega}=\sqrt{\frac{s'/s}{\varkappa \varkappa'}}(cs)^{-1/3}
e^{i(b/c)(s-s')-i(2/3)w^{3/2}}\,[\pi\,{\rm Ai}(z)-i(\pi\,{\rm Gi}(z)-\frac 1z)],
\end{equation}

\begin{equation}
\alpha^F_{\omega'\omega}=\frac{(cs)^{-1/3}}{\sqrt{\varkappa\varkappa'
(\alpha +1)}}\,e^{i(b/c)(s-s')-i(2/3)w^{3/2}}[\frac{i\sqrt{\alpha}}{z}+
\frac{1}{\sqrt{w}}\int_0^\infty dt\,\sqrt{t^2+\alpha w}\,e^{-izt-it^3/3}].
\end{equation}
Here ${\rm Ai}(z)$ and ${\rm Gi}(z)$ are well known Airy and Scorer functions 
defined as in [28], and
$$
z=(cs)^{-1/3}(s-s')-w,\quad w=(b/c)^2(cs)^{2/3},\quad s=\omega/\varkappa,\quad
s'=\omega'/\varkappa',\quad \alpha=c/b^2-1,
$$
Parameter $\alpha = (1-\beta_{10})/2\beta_{10}$ is defined by the initial 
relative velocity $\beta_{10}$ of the mirror and detector, 
$\beta_{20}=-1$.

At the finding ${\rm tr}\,\alpha^B$ the integral
\begin{equation}
{\rm tr}\,\alpha^B=\frac{1}{2\pi}\,\frac 32(\alpha+1)\int_0^\infty dw\,
e^{-i\frac 23 w^{3/2}}[\pi{\rm Ai}(-w)-i(\pi{\rm Gi}(-w)+\frac {1}{w})]
\end{equation}
appears which diverges both on the lower and upper limits. The infrared
divergency is remedied by introducing the nonzero lower limit $w_1=
(\varepsilon/2(\alpha+1)^2)^{2/3}$ where $\varepsilon = \nu/w_0 \ll 1$. To 
eliminate the ultraviolet divergency we subtract from the integrand the first 
term $\sqrt{\pi} e^{-i\pi/4} w^{-1/4}$ its asymptotical expansion for 
$w \to \infty$. Now it is possible to turn the integration contour on the angle 
$-\pi/3$ and, introducing the integration variable $t=e^{i\pi/3} w$, to brought 
the regularized integral to the form
\begin{equation}
{\rm tr}\,\alpha^B\vert_{reg}=\frac{1}{2\pi}\,\frac32 (\alpha+1)\lbrace 
-\frac{\pi}{3}-i\int_{t_1}^\infty \frac{dt}{t}\,e^{-\frac23 t^{3/2}} +
i\int_0^\infty dt\,\pi {\rm Gi}(t)e^{-\frac23 t^{3/2}}\rbrace .
\end{equation}
By the way we used the formulae
\begin{gather}
{\rm Ai}(e^{2\pi i/3} t)=\frac12 e^{i\pi/3}[{\rm Ai}(t)-i{\rm Bi}(t)],\quad
{\rm Gi}(e^{2\pi i/3} t)=-e^{i\pi/3}{\rm Gi}(t)+\frac12 e^{-i\pi/6}[{\rm Ai}(t)+
i{\rm Bi}(t)], \notag\\
\int_0^\infty dt\,(\pi{\rm Bi}(t)e^{-\frac23 t^{3/2}}-\frac{\sqrt{\pi}}{t^{1/4}})
=0.
\end{gather}
The last integral in (118) is equal to $\frac23 +\frac29 \ln 2$. As a result we 
obtain finally
\begin{equation}
{\rm tr}\,\alpha^B\vert_{reg} =\frac{1}{2\pi}(\alpha +1)\lbrace -\frac{\pi}{2}-
i[\ln\,\frac{3(\alpha + 1)^2}{\gamma\varepsilon}-1-\frac13\,\ln 2]\rbrace,\quad 
\varepsilon = \nu/w_0,
\end{equation}

The evaluation of ${\rm tr}\,\alpha^F$ follows by the similar way. Now the 
integral
\begin{equation}
{\rm tr}\,\alpha^F=\frac{1}{2\pi}\,\frac 32\sqrt{\alpha+1}\int_0^\infty dw\,
e^{-i\frac 23 w^{3/2}}[\frac{1}{\sqrt{w}}\int_0^\infty dt\,\sqrt{t^2+\alpha w} 
\,e^{iwt-it^3/3} -\frac {i}{w}\sqrt{\alpha}\,]
\end{equation}
appears instead of the integral (117). The main terms of asymptotical expansions
of the integrand for $w \to 0$ and $w \to \infty$ are identical to those of the 
integrand in (117) and differ from them only by extra multipliers $\sqrt{\alpha}$ 
and $\sqrt{\alpha+1}$ correspondingly. After elimination of the infrared and 
ultraviolet divergences and turning the integration contour on the angle 
$-\pi/3$ we obtain

\begin{equation}
{\rm tr}\,\alpha^F\vert_{reg} = \frac{1}{2\pi}\lbrace \sqrt{\alpha(\alpha+1)}(-\frac{\pi}
{2}-i\ln\frac{3(\alpha+1)^2}{\gamma\varepsilon}) + i\sqrt{\alpha+1}\,J(\alpha)
\rbrace,
\end{equation}
where
\begin{equation}
J(\alpha)=-3\int_0^\infty dx\,[\,e^{-\frac23 x^3}\int_0^\infty d\tau\,\sqrt{\tau^2
+\alpha x^2}\,e^{x^2\tau - \tau^3/3}-\sqrt{\pi(1+\alpha)x}\,].
\end{equation}
S.L. Lebedev called author's attention to the fact that the integral 
$J(\alpha)$ can be reduced to elementary functions. Indeed, it can be shown that
\begin{equation}
J(\alpha)=1+\sqrt{\alpha}+\frac{\alpha -2}{3\sqrt{\alpha +1}}\ln\frac{\alpha+
\sqrt{\alpha (\alpha+1)}}{1+\sqrt{\alpha+1}}+\frac{\sqrt{4+\alpha}}{3}\ln\frac
{\sqrt{\alpha (4+\alpha)}-\alpha}{4+2\sqrt{4+\alpha}}.
\end{equation}
The function $J(\alpha)$ is equal to $1-\frac23 \ln\,2=0.5379\ldots\,$ at 
$\alpha=0$, attains minimal value $\approx 0.39$ at $\alpha \approx 0.3$, and 
then grows and behaves as $(1+\frac13 \ln\,2)\sqrt{\alpha}$ as $\alpha \to 
\infty$.

Note, that $\alpha^{B,F}_{\omega'\omega}$ depend on two dimensionless parameters
$b,\:c$, but the traces ${\rm tr}\,\alpha^{B,F}$ depend only on their 
combination $\alpha$, i.e. only on the subluminal relative velocity 
$\beta_{10}$.
                
Airy semiparabola with out-tangent line is time-reversed to the considered
trajectory  and can be obtained from it by the changes $v\rightleftarrows -u,\;
\varkappa \rightleftarrows \varkappa'$. This leads to the change 
$s\rightleftarrows s'$ in the expressions for $\alpha^{B,F}_{\omega'\omega}$.
The ${\rm tr}\,\alpha^{B,F}$ do not change at all, but it must be understood
that the parameter $\alpha$ is now defined by the final (and negative) relative
velocity $\beta_{20}$ of the mirror and detector: $\alpha=-(1+\beta_{20})/
2\beta_{20} >0$, while $\beta_{10}=1$.

The infrared logarithmic singularities of ${\rm tr}\,\alpha^{B,F}$ were 
regularized by nonzero momentum transfer $\nu \ll w_0$. Their coefficients are 
in accordance with general consideration of Section 6. These singularities
disappear from ${\rm tr}\,\alpha^F\vert_{reg}$ at luminal velocities of the 
ends, and ${\rm tr}\,\alpha^F\vert_{reg}$ becomes pure imaginery positive.
The positive sign of the Im\,${\rm tr}\,\alpha^F\vert_{reg}$ in this case
can be explained by the large momentum transfer to the mirror during its 
contact with detector while the negative signs of Im\,$\varDelta m_0$ and
Im\,$\varDelta m_1$ are connected with energy-momentum losses by a charge due 
to the change of self-interaction at acceleration.

We do not consider here the coefficients $\beta^{B,F*}_{\omega'\omega}$. They
can be obtained from $\alpha^{B,F}_{\omega'\omega}$ by the changes $\omega \to
-\omega,\:\sqrt{\omega} \to -i\sqrt{\omega}$, and division on $i$ in Bose-case,
see (2).
                                                
\section{Conclusion}
The symmetry being discussed was revealed itself in the coincidence of the 
bilinear in $\beta_{\omega'\omega}$ quantities, such as
$$
\vert\beta_{\omega'\omega}\vert^2,\quad (\beta^+\beta)_{\omega\omega}=\int_0
^\infty \frac{d\omega'}{2\pi}\,\beta^*_{\omega'\omega}\,\beta_{\omega'\omega},
\quad \bar N={\rm tr}\,\beta^+\beta=\int_0^\infty \frac{d\omega}{2\pi}\,
(\beta^+\beta)_{\omega\omega},
$$
with the corresponding quantities describing the emission of vector (scalar)
quanta by electric (scalar) charge in 3+1-space, see Introduction. Only likely
transforming frequencies are involved in each summation entering in these 
quantities as well as in the equality $\omega=\omega''$ for the diagonal
elements of the matrix $(\beta^+\beta)_{\omega\omega''}=\int_0^\infty
\frac{d\omega'}{2\pi}\,\beta^*_{\omega'\omega}\beta_{\omega'\omega''}$. On the
other hand, the definition of the trace of matrix $\alpha_{\omega'\omega}$
with differently transforming indexes $\omega,\,\omega'$ required the Lorentz-
invariant frequencies $\Omega,\,\Omega'$, coinciding with $\omega,\,\omega'$ in
the proper system of the detector, moving along the tangent line to the mirror's
trajectory at characteristic point. As a result, the ${\rm tr}\,\alpha$ becomes
a functional of not only the mirror's trajectory but also the detector's one.
This allows to consider the ${\rm tr}\,\alpha$ as experimentally measurable
quantity.

The symmetry under discussion has been embodied in several exact 
mathematical relations between important physical quantities. The most important
of them are, of course, the fundamental relations (11), (12) between the 
Bogoliubov coefficients for the processes induced by a mirror in 1+1-space
and the current and charge densities for the processes induced by a charge
in 3+1-space. Another one is the integral connection (16) between the propagator 
of a pair of massless particles, scattered in 1+1-space in opposite directions 
with frequencies $\omega,\,\omega'$ (so that the pair has a mass 
$m=2\sqrt{\omega \omega'}$), and the propagator of a single particle in 
3+1-space. This relation provides the connection $\varDelta W_{1,0}=
e^2\,\varDelta W^{B,F}$ between the self-action changes of a charge in 
3+1-space and of a mirror in 1+1-space if ${\rm tr}\,\beta^+\beta \ll 1$.

The other relations with the symmetry embodied in are the spectral 
representations for the real parts of the self-action changes (32) and of the
mass shifts (34),(38) of electric and scalar charges in quasihyperbolic motion.
So, the mass shifts of a charges, the sources of the Bose-fields with spin 1
and 0 in 3+1-space, are represented by the spectral distributions of the heat
capacity or the energy of Bose- and Fermi-gases of massless particles in 
1+1-space. The spectral representations allow to consider the mass shift
formation as the balance between the radiation and excitation of the proper 
energy at acceleration.
       
The symmetry between processes induced by the mirror in two-dimensional and by 
the charge in four-dimensional space-times predicts not only the value
$e_0^2=1$ for the bare charge squared that corresponds to the bare fine 
structure constant $\alpha_0=1/4\pi$. It predicts also the appearance of scalar 
particles in ultra high-energy collisions in 3+1-space and the decreasing their 
interaction with scalar source with increasing of the energy.

It is very interesting that the bare fine structure constant has the purely 
geometrical origin, and, also, that its value is small: $\alpha_0=1/4\pi\ll 1$.
The smallness of $\alpha_0$ has the essential meaning for the quantum 
electrodynamics where it explains the smallness of $\alpha$ and justifies 
a priori the applicability of the perturbation theory.                                                       

I am grateful to M.A. Vasiliev for useful discussions and comments.                                  

The work was carried out with financial support of Scientific Schools and
Russian Fund for Fundamental Research (Grants 1578.2003.2 and 05-02-17217).

\end{document}